\documentclass[10pt,journal]{IEEEtran}

\usepackage{amssymb,upgreek}
\usepackage[final]{graphicx}
\usepackage{amsmath}
\usepackage{amsfonts,bm}
\usepackage{mathtools,color}
\usepackage[caption=false,font=footnotesize]{subfig} 
\usepackage{multirow,booktabs}
\usepackage{cite}

\usepackage{epstopdf}
\graphicspath{{eps/}}

\newcommand{\review}[1]{\textcolor{black}{#1}}

\newcommand{\C}{ \mathbb{C}}
\newcommand{\R}{ \mathbb{R}}
\newcommand{\Z}{ \mathbb{Z}}
\newcommand{\N}{ \mathbb{N}}

\def\SNR{\mathrm{SNR}}

\newcommand{\Lop}{{\rm L}}

\newcommand{\ee}{\mathrm{e}}

\newcommand{\ue}{\mathrm{e}}
\newcommand{\uj}{\mathrm{j}}

\newtheorem{definition}{Definition}
\newtheorem{proposition}{Proposition}

\newtheorem{theorem}{Theorem}

\def\ee{\mathrm{e}}     

\def\Var{\mathrm{Var}}

\DeclareMathOperator*{\argmax}{arg\,max}

\begin{document}

\title{Principled Design and Implementation \\ of Steerable Detectors}

\author{Julien Fageot$^{\ast}$, Virginie Uhlmann$^{\ast}$, Zsuzsanna P\"usp\"oki, Benjamin Beck, Michael Unser, Adrien Depeursinge
\thanks{This project was supported by the Swiss National Science Foundation under Grants 200020-162343/1, PZ00P2\_154891, and 205320\_179069, and by EMBL internal funding.}
\thanks{Julien Fageot is with the Biomedical Imaging Group, EPFL, 1007 Lausanne, Switzerland, and with the Signals, Information, and Networks Group, Harvard University, Cambridge, MA 02138, USA (e-mail: julien.fageot@gmail.com).}
\thanks{Virginie Uhlmann is with the European Bioinformatics Institute (EMBL-EBI), Cambridge CB10 1SD, UK (e-mail: uhlmann@ebi.ac.uk)}
\thanks{Zsuzsanna P\"usp\"oki, Benjamin Beck, and Michael Unser are with the Biomedical Imaging Group, EPFL, 1007 Lausanne, Switzerland.}
\thanks{Adrien Depeursinge is with the Biomedical Imaging Group, EPFL, 1007 Lausanne, Switzerland, and with the MedGIFT group, Institute of Information Systems, University of Applied Sciences Western Switzerland (HES-SO), 3960 Sierre, Switzerland.

$^\ast$These two authors contributed equally to this work.}}

\maketitle

\begin{abstract}
We provide a complete pipeline for the detection of patterns of interest in an image.
In our approach, the patterns are assumed to be adequately modeled by a known template, {and are located at unknown positions and orientations that we aim at retrieving.}
We propose a continuous-domain additive image model, where the analyzed image is the sum of the patterns to localize and a {background  with self-similar isotropic power-spectrum}. 
We are then able to compute the optimal filter fulfilling the SNR criterion based on one single template and background pair: it strongly responds to the template while being optimally decoupled from the background model.
{In addition, we constrain our filter to be steerable, which allows for a fast template detection together with orientation estimation.}
In practice, the implementation requires to discretize a continuous-domain formulation on polar grids, which is performed using quadratic radial B-splines. 
We demonstrate the practical usefulness of our method on a variety of template approximation and pattern detection experiments.
{We show that the detection performance drastically improves when we exploit the statistics of the background via its power-spectrum decay, \review{which we refer to as spectral-shaping}.
The proposed scheme outperforms state-of-the-art steerable methods by up to 50\% of absolute detection performance.}
\end{abstract}

	\vspace{-0.3cm}	
\section{Introduction} \label{sec:intro}

{ Pattern detection and recognition is a core task of image analysis in general~\cite{sonka2014image}, and finds applications in physics, chemistry, medicine, and biology~\cite{Ran2004}. In images at the microscopic scale, patterns of interest are characterized by pronounced and characteristic directional components~\cite{frangi1998multiscale,depeursinge2017multiscale,depeursinge2018rotation,PTB2009}
and well-defined geometrical structures, such as centrioles~\cite{Guichard2013}, nuclear pores~\cite{Beck2004}, and flagella bodies~\cite{Kawamoto2013} in biology, or foam junctions~\cite{Mader2012} and crystals in physics.
There, structures of interest do not only appear at different locations in the image, but also at various orientations.}
It follows that an important challenge for accurate pattern detection is to develop detectors that can sense discriminative directions with invariance or equivariance to translations and local rotations~\cite{Depeursinge2018biomedical}.


\vspace{-0.3cm}
{\subsection{The Detection Problem at a Glance}

We focus on the detection of structured patterns composed of repetitions of a motif with few variations. The pattern can thus adequately  be  modeled by a single template instance. }
This leads us to formulate the detection problem as follows.
We assume that the image $I$  to analyze is such that
\begin{equation} \label{eq:detectionImageModelintro}
	{I} = \sum_{j=1}^J {T} ( \mathbf{R}_{- \theta_j} ( \cdot  - \boldsymbol{x}_j ) )+ {S},
\end{equation}
	with $T$ a known template dispatched at $J$ unknown locations $\bm{x}_j$ and rotated at unknown orientations $\theta_j$, and $S$ some background, modeled as an isotropic and self-similar Gaussian random field.
	
	Our approach is to build a filter that allows for a single-shot convolution-based detection through the complete image, carried out in the Fourier domain.
	We use steerable filters to efficiently test any possible orientation at any image position, without the need for reconvolving the image with pre-computed rotated versions of the detector as would be done by classical rotated template matching.
	The detector is designed to be discriminative in the sense that it responds strongly to the template $T$, while being as insensitive as possible to the background $S$.  {\color{black}The latter is achieved by adapting the radial Fourier transform of the steerable detector to the power-spectrum of $S$, an operation we refer to as \emph{spectral shaping}.}
	
	An important design feature of our method is that the detection filter can be obtained from one \emph{unique} example (the template $T$ provided by the user) and from the background model (characterized by a single parameter quantifying its power-spectrum decay, as we shall describe later).
	{ We demonstrate that it is possible to adapt the steerable detector to the template and the background's second order statistics in a very efficient way with excellent approximation performance, and then use the steerability property to enable fast and accurate pattern detection and orientation estimation (parameters $\bm{x}_j$ and $\theta_j$ in \eqref{eq:detectionImageModelintro}, respectively).}
	\review{Since we provide a principled and efficient methodology for a general detection task, our method could easily be incorporated in a similar manner as template matching into more involved processing pipelines for, \emph{e.g.}, electron microscopy image analysis~\cite{Schorb2019software, yin2020petascale} or manufacturing~\cite{kuo2019automatic}.}
	
\vspace{-0.3cm}
\subsection{Contributions}
	
	Our main contributions are summarized as follows.

{ \emph{Theory: {\color{black}Optimal steerable detector with spectral shaping.}} We define an SNR criterion for the image model \eqref{eq:detectionImageModelintro}, from which we derive the optimal steerable filter for pattern detection. When the background has a flat power-spectrum (\emph{i.e.}, white noise model), the filter is the steerable function that best matches the template $T$ (Theorem \ref{theo:optimalfilter}). 
\review{One of our main contributions is to} consider a richer background model with self-similar and isotropic power-spectrum, for which we also derive the optimal steerable filter (Proposition \ref{prop:spectralshaping}). It consists of a spectrally shaped version of the optimal steerable filter for the flat-spectrum model.

 \emph{Implementation: Radial B-spline expansion.} The optimal steerable filter is characterized by its Fourier domain angular/radial decomposition.
		The angular dependency is controlled by the use of circular harmonics, which yields an angular low-pass approximation of the optimal filter.
		The radial profile of the optimal filter is captured by developing an interpolation method based on radial B-splines, relying on the identification of the optimal B-spline-based steerable filter in Theorem \ref{theo:Bsplineexpansion}.
		 We are therefore able to construct, from a single occurrence of a template and background model, the spectrally shaped spline-based steerable filter that maximises the SNR criterion.
		
	 \emph{Template detection pipeline.} We provide a complete template detection algorithm that we evaluate experimentally.  We demonstrate the effectiveness of our approach in comparison to steerable filters built from Hermite kernels~\cite{yang2013steerability}  \review{and from the closely-related Fourier-Argand representation~\cite{zhao2020fourier}. 
	Our method clearly outperform competing approaches thanks to the ability of our optimal detector to adapt to the background model (spectral shaping).}
	}

	\vspace{-0.3cm}	
\subsection{Related Works}
A whole range of classical image processing methods for pattern detection are based on handcrafted filters, transforms and criteria (\emph{e.g.}, Hough~\cite{Bal1981}, Laplacians of Gaussians~\cite{MaH1980}, Canny~\cite{Can1986}, Harris~\cite{HaS1988}). They are targeting low-level image features such as lines, blobs, edges, lines or corners and do not allow modeling more complex patterns in arbitrary templates.

Two main types of approaches are available to detect possibly rotated versions of an arbitrary template: template matching, and steerable filters.
Template matching allows locating occurrences of virtually any object that can be modeled by a template by minimizing a given matching distance between the template and a local image neighborhood~\cite{OTM2012} (\emph{e.g.}, sum of absolute differences, normalized cross-correlation).
However, rotationally invariant versions of template matching methods are computationally intensive since matching distances must be evaluated for every positions and orientations of the pattern.
{ Moreover, standard template matching algorithms are---often implicitly---designed for backgrounds with flat power spectrum and therefore require data to be pre-processed, for instance by applying local background suppression (LBS).}


{ Steerable filters offers an efficient alternative to detect the locations and orientations of a pattern's occurrence in an image~\cite{Simoncelli1995steerable,Liu2012,muehlich2012design,yang2013steerability,Puspoki2015template,puspoki2016design,yang2017recognition}.
The steerability property allows evaluating a filter's response at any real-valued orientation by using a simple angular-dependent linear combination of a small number of basis elements~\cite{Freeman1991design,Unser2013steerable}, which yields computationally fast algorithms for position detection~\cite{puspoki2016design,yang2017recognition,zhao2020fourier} and orientation estimation~\cite{puspoki2019angular}.
While they are mostly known for their ability to detect ridges and edges, steerable filters can be shaped to provide a low-pass approximation of any generic template. 
\review{State-of-the-art steerable-based methods for pattern detection rely on Hermite kernels~\cite{yang2013steerability} or on the Fourier-Argand representation of steerable filters~\cite{zhao2020fourier}. 
As such, these two approaches are the most closely related to our work and will therefore be our main points of comparison.}


The approach we propose has the specificity to be adaptive: our framework builds an optimal steerable approximation of any template of interest relying on a custom spline-based framework. It considerably reduces the number of basis function needed to  adequately represent  the template, improving the detection performances. \review{A similar principle was recently exploited in the closely-related Fourier-Argand representation~\cite{zhao2020fourier}.
Our method however includes in addition a background model in the filter design criteria, which spectrally shapes the optimal detector in order to maximally differentiate it from the background signal. 
To the best of our knowledge, this strategy has never been proposed in the steerable literature so far. It dramatically improves detection performance in practice and thus offers a theoretically sound and more sophisticated alternative to standard pre-processing strategies required by template matching.}


{We deem it essential to position our work with respect to deep convolutional neural networks (CNN) approaches considering the tremendous progress they have allowed for pattern detection in recent years~\cite{RIZ2017,SRG2016}. 
CNNs are able to learn a collection of detectors as deep image operators that are invariant to translations via convolutional operations. In their initial formulation, CNNs have no built-in invariance to rotations, which is often palliated using rotational data augmentation~\cite{simard2003best}.
More advanced and recent approaches, including group equivariant and steerable CNNs~\cite{weiler2018learning,sifre2013rotation,WGT2016,CoW2016b,bekkers2018roto,cohen2019general,andrearczyk2019exploring,andrearczyk2020local,oreiller20203d}, aim at bridging this gap by designing rotation invariant network architectures.
Nevertheless, CNNs require a large number of training samples of object and background to adequately learn deep image detectors, and  
cannot be trained from a single example.
The problem tackled in this paper, in comparison, is to build a directly interpretable detector from a \emph{single} occurrence of a template, an application for which CNNs are clearly unsuitable.}


\vspace{-0.3cm}	
	\subsection{Outline} 

The paper is structured as follows. The continuous-domain theory of SNR-based optimal steerable filter design 
is presented in Section~\ref{sec:theory}.
The implementation of the proposed theory on discretized image grids and detection algorithm are detailed in Section~\ref{sec:OSFimplementation}.
The performance and parameter sensitivity of the proposed framework is investigated in  Section~\ref{sec:OSFpracticing}.
Finally, discussions and conclusions are presented in Section~\ref{sec:discussandconclude}.

\vspace{-0.3cm}

\section{Optimal Steerable Filters: Theory} \label{sec:theory}

	This section is dedicated to our continuous-domain framework for the detection problem. 
	After introducing the main notations in Section \ref{subsec:notations}, 
	we present the SNR criterion for which the optimal detectors will be constructed in Section \ref{subsec:SNR}. 
	One challenging aspect of the detection is that the pattern can be found at an unknown orientation. We address this problem using steerable filters, introduced in Section \ref{subsec:steer}. 
	The main theoretical result of this paper is Theorem \ref{theo:optimalfilter} in Section \ref{subsec:optimalfilter}, which gives the formula for the optimal steerable detector considering the SNR criterion. 
	In Section \ref{subsec:spectralshaping}, we present a refinement of this result for a richer class of background models that appears to be much more realistic than the flat spectrum model. 
	Finally, we show how to define response maps based on the optimal detector, that can be used for the detection procedure in Section~\ref{sec:detectionprocedure}. 

\vspace{-0.3cm}
	\subsection{Notations} \label{subsec:notations}

	Vectors in the plane are denoted in spatial domain by $\bm{x} = (x_1,x_2) \in \R^2$  and in Fourier domain by $\bm{\omega} = (\omega_1, \omega_2)\in \R^2$.
We write $(r,\theta)$ for the polar coordinates in Fourier domain  where $r\geq 0$ and $\theta \in [0,2\pi)$.
We switch from Cartesian to polar coordinates according to $(\omega_1,\omega_2) = (r \cos\theta, r\sin\theta)$ and $(r,\theta)= \left( (\omega_1^2+\omega_2^2)^{1/2}, \arctan(\omega_2/ \omega_1)\right)$.

We consider functions $f$ from $\R^2$ to $\R$.  The Fourier transform of $f$ is $\widehat{f}$. A function $f$ is square integrable and denoted by $f\in L_2(\R^2)$  if $\lVert f\rVert_2^2 = \int_{\R^2} \lvert f(\bm{x})\rvert^2\mathrm{d} \bm{x} < \infty$. 
We shall repeatedly use the Parseval relation $\langle f , g \rangle = \frac{1}{2\pi} \langle \widehat{f} , \widehat{g} \rangle$. 
The usual scalar product between two square integrable functions in Fourier domain is then
\begin{align*}
	\langle \widehat{f} , \widehat{g} \rangle & =  \int_{\R^2} \widehat{f}(\bm{\omega}) \overline{\widehat{g}(\bm{\omega})} \mathrm{d}\bm{\omega} =  \int_{0}^{2\pi}  \int_{0}^\infty \widehat{f}(r,\theta) \overline{\widehat{g}(r,\theta)} r \mathrm{d} r \mathrm{d} \theta.
\end{align*}
The rotation matrix of angle $\alpha$ is $\mathbf{R}_{\alpha} = \begin{pmatrix} 
\cos \alpha & - \sin\alpha \\
\sin \alpha & \cos \alpha 
\end{pmatrix}$.
Finally,  $f\propto g$ means that the two functions $f$ and $g$ are proportional, \emph{i.e.}, $f = \lambda g$ with $\lambda \neq 0$. 

\vspace{-0.3cm}
	\subsection{Local Image Model and SNR Criterion} \label{subsec:SNR}

{We consider the local version of the image model \eqref{eq:detectionImageModelintro}, where a single template is located at the center of the image. In this local image model,  the image $I_{0}$ is the sum of a template of interest $T$ and a background $S$ as}
\begin{equation} \label{eq:imageModel}
	{I}_0 (\bm{x}) = {T} (    \bm{x}   )+ {S} (\bm{x}).
\end{equation}
Mathematically, $T$ is a square-integrable function and $S$ is modeled as a Gaussian field with zero mean.
The goal is to design detection filters $f$ that (i) strongly responds to the foreground template ${T}$, (ii) responds as little as possible to the background ${S}$, and (iii) can be used efficiently to determine the orientation of the template ${T}$ when it is unknown.

The third requirement will be achieved by using steerable filters, introduced in Section \ref{subsec:steer}.
To tackle the two first point, we want $f$ to maximizes the \textit{signal-to-noise ratio} (SNR), defined as
\begin{equation} \label{eq:SNR}
	\SNR(f) = \frac{\mathbb{E} [\langle I_0 , f \rangle ]^2}{\mathrm{Var} (\langle I_0, f \rangle)},
\end{equation}
{with $\mathbb{E}$ the expected value and $\mathrm{Var}$ the variance of the considered random variables.}
The SNR criterion is a classical metric in detection theory \cite{Van2004}.
The template ${T}$ is deterministic, hence we have that $\mathbb{E} [\langle I_0 , f \rangle ] = \langle T , f\rangle$ and $\Var( \langle I_0 , f \rangle ) = \Var( \langle S , f\rangle)  $. 

\vspace{-0.3cm}
	\subsection{Steerable Filters and their Fourier Radial Profiles} \label{subsec:steer}

	We aim at detecting patterns whose orientations are \emph{a priori} unknown in an image.
	This can be performed using steerable filters, which can be rotated efficiently~\cite{Unser2013steerable}. \\
	\vspace{-0.3cm}
	\begin{definition} \label{def:steerable}
A filter $f$ is \emph{steerable} if the span of its rotated versions $f(\mathbf{R}_{\alpha}\cdot)$, with $\alpha\in [0,2\pi)$, is a finite-dimensional subspace of $L_2(\R^2)$. {This means that there exist $P \geq 1$ filters $f_p$, $p=1\ldots P$ such that
, for any rotation angle $\alpha \in [0,2\pi)$, we have  
$f (\mathbf{R}_{\alpha} \cdot ) = \sum_{p=1}^P c_p(\alpha) f_p,$
for some  $c_p(\alpha) \in \R$, $1 \leq p \leq P$.} \\
\end{definition} 
\vspace{-0.3cm}

The main advantage of steerable filters is that their rotation by an arbitrary angle is reduced to a finite dimensional algebraic problem, { allowing for fast pattern detection even when the orientation is unknown.}
In Proposition \ref{theo:harmonicdecomposition}, we characterize steerable filters from their polar decomposition in terms of the circular harmonic (CH) functions $\theta \mapsto \mathrm{e}^{\mathrm{j} n \theta}$, where $n\in \Z$.  \\ 

\vspace{-0.3cm}
\begin{proposition} \label{theo:harmonicdecomposition}
	A function $f \in L_2(\R^2)$ can be uniquely decomposed in polar coordinates in Fourier domain as
	\begin{equation} \label{eq:ladecomposition}
		\widehat{f}(r,\theta) = \sum_{n\in \Z} \widehat{f}_n(r) \ee^{\uj n \theta},
	\end{equation}
	where the $\widehat{f}_n$ jointly satisfy
$	\sum_{n\in \Z} \lVert \widehat{f}_n\rVert_2^2 < \infty.$	
The functions $\widehat{f}_n$, called the \emph{Fourier radial profiles} of $f$, are given by
	\begin{equation} \label{eq:fn}
		\widehat{f}_n(r) = \frac{1}{2\pi} \int_0^{2\pi} \widehat{f}(r,\theta) \ee^{- \uj n \theta} {\rm d} \theta.
	\end{equation}
		Moreover, $f$ is steerable if and only if finitely many $\widehat{f}_n$ are non-zero. \\
\end{proposition}
\vspace{-0.3cm}

\review{Proposition~\ref{theo:harmonicdecomposition} was also enunciated  in~\cite[Section II-B]{zhao2020fourier}, in which \eqref{eq:ladecomposition} is referred to as the \emph{Fourier-Argand representation}. No proof was however provided, and we therefore include one in Appendix \ref{ProofDecomposition} for the interested reader.}
As a consequence, the general form of a steerable filter in the Fourier domain is
	$\widehat{f} (r,\theta) = \sum_{n\in H} \widehat{f}_n (r) \mathrm{e}^{\mathrm{j}n \theta}$,
where $H$ is a finite subset of $\Z$ and $\widehat{f}_n \in L_2(\R^2)$ the non zero Fourier radial profiles. We then have
\begin{equation} \label{eq:fsteerableRotation}
\widehat{f (\mathbf{R}_{\alpha} \cdot)} (\bm{\omega}) =	\widehat{f}(\mathbf{R}_{\alpha} \boldsymbol{\omega}) = \widehat{f}(r, \theta + \alpha) = \sum_{n \in H}  \mathrm{e}^{\mathrm{j} n \alpha} \widehat{f}_n(r) \ue^{\uj n \theta}.
\end{equation}
Hence, any rotated version of $f$ is a linear combination of the inverse Fourier transforms of the $\widehat{f}_n(r) \ue^{\uj n \theta}$ for $n\in H$, meaning that $f$ is steerable in the sense of Definition \ref{def:steerable}.

Finally, any function $f$ can be approximated by steerable functions at an arbitrary precision. Indeed, it is sufficient to consider the truncated sums $\sum_{\lvert n \rvert \leq N} \widehat{f}_n(r) \ue^{\uj n \theta}$ that converge to $\widehat{f}$ in  $L_2(\R^2)$ when the number of harmonics $N$ increases.
 
 \vspace{-0.3cm}
	\subsection{Optimal Steerable Filter Design for White Background} \label{subsec:optimalfilter}

In this section, we assume that the background $S$ is a Gaussian white noise, which corresponds to a flat power spectrum. 
This implies that $ \Var( \langle S , f\rangle) =  \sigma^2 \lVert f \rVert_2^2$, with $\sigma^2$ the variance of $S$ (see Appendix \ref{app:randomModels}).
As a consequence, \eqref{eq:SNR} becomes
\begin{equation}\label{eq:SNR2}
	\SNR(f) = \frac{1}{\sigma^2} \frac{ \lvert \langle {T} , f \rangle\rvert^2}{ \lVert f \rVert_2^2} = \frac{1}{\sigma^2} \frac{ \lvert\langle \widehat{T} , \widehat{f} \rangle\rvert^2}{ \lVert \widehat{f} \rVert_2^2},
\end{equation}
where we used the Parseval relation for the Fourier domain expression.
In Section \ref{subsec:spectralshaping}, we will also consider more evolved background models.

For a given finite set of harmonics $H$, it is then possible to specify the optimal steerable filter for the SNR criterion \eqref{eq:SNR2} associated to the image model \eqref{eq:imageModel}. \\
\vspace{-0.3cm}
\begin{theorem} \label{theo:optimalfilter}
	A filter $f$ maximizes the SNR criterion \eqref{eq:SNR2} among the space of steerable filters with harmonics in $H$ if and only if 	
	\begin{equation} \label{eq:fSNR}
		\widehat{f}(r,\theta) \propto \sum_{n \in H} \widehat{T}_n (r) \mathrm{e}^{\mathrm{j} n \theta},
	\end{equation}
	with $\widehat{T}_n$   the Fourier radial profiles of $T$ given for each harmonic $n$ by
	\begin{equation} \label{eq:integralTn}
		\widehat{T}_n(r) = \frac{1}{2\pi} \int_{0}^{2\pi} \widehat{T}(r,\theta) \mathrm{e}^{-\mathrm{j} n \theta} \mathrm{d} \theta.
	\end{equation}
\end{theorem}
 
 The optimal filter is defined up to a multiplicative constant since $\SNR(\lambda f) = \SNR (f)$ for every scalar $\lambda \neq 0$.
The proof of Theorem \ref{theo:optimalfilter} is given in Appendix \ref{ProofOptiFilter}.
The optimal filter is completely determined by the template to approximate $T$ and the set of harmonics $H$.
In practice, the main issue is to compute the integral \eqref{eq:integralTn} while knowing only $T$ on a finite cartesian grid in the Fourier domain.
This point will be discussed extensively in Section \ref{sec:OSFimplementation}.

\vspace{-0.3cm}
	\begin{tiny}
	\subsection{Isotropic Self-similar Background  and Spectral Shaping} \label{subsec:spectralshaping}
	\end{tiny} 
	
	The SNR criterion \eqref{eq:SNR2} is based on the assumption that the background $S$ in \eqref{eq:detectionImageModelintro} is adequately modeled as a Gaussian white noise, corresponding to a nearly constant power spectrum $P_S(\bm{\omega})$.
	{We  introduce the basics to determine the SNR criterion for a richer statistical background model. Additional mathematical details are provided in Appendix \ref{app:randomModels} for the interested reader.}
	
	It has been shown in many signal and image processing applications that the power spectrum of the signal of interest follows a power law~\cite{Puspoki2015template,Flandrin1992wavelet,Sage2005automatic,Unser2014sparse},
	and is therefore smoother than a white noise.
	We shall adopt such a model here for the background  noise $S$, while further assuming that it  is statistically isotropic, which is equivalent to saying that the power spectrum is a radial function $P_S(\bm{\omega})=P_S(r)$ with $r = \lVert \bm{\omega} \rVert$.
	{Mathematically, this means that the power spectrum of the background $S$ is equal to $P_S(\bm{\omega}) = P_S(r) = \sigma^2 / r^{2\gamma}$, where $\sigma^2$ is the variance. 
	
	Such a background $S$ is called an \emph{isotropic self-similar (ISS) Gaussian field}~\cite{Tafti2010brownian,Lodhia2016fractional,FageotThesis}, and
	$\gamma$ is its \emph{self-similarity parameter}. For illustration purposes, we represent several ISS Gaussian fields in Fig.~\ref{fig:seeprocess}, where the theoretical fact that higher $\gamma$ yield smoother $S$~\cite[Corollary 1]{fageot2020n} can be visually observed. }
	
\begin{figure}[t!]\label{fig:IM}
\centering
\includegraphics[scale=0.30]{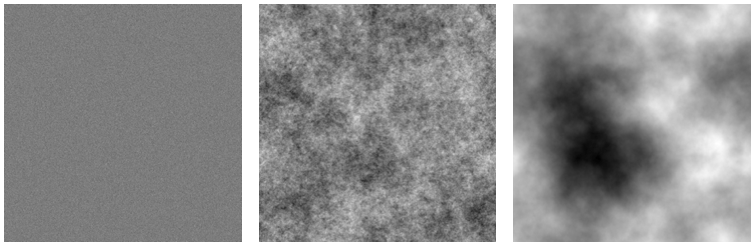}
\caption{Realizations of ISS Gaussian fields for different values of $\gamma$. From left to right: $\gamma = 0$ (white noise), $\gamma = 1$, $\gamma=2$.}
\label{fig:seeprocess}
\end{figure}
	
	From Appendix \ref{app:randomModels}, we deduce the general implications of this background model.	
	First, we shall consider filters such that $\langle S , f\rangle$ is well-defined, which requires that $\widehat{f}(\bm{\omega})   / \lVert \bm{\omega} \rVert^{\gamma} \in L_2(\R^2)$. 
Then, $\langle S, f \rangle$ is a well-defined Gaussian random variable with zero mean and variance 
	\begin{align} \label{eq:varianceS}
		\mathrm{Var}(\langle S , f \rangle) &=   \frac{\sigma^2 }{2\pi} \int_0^{\infty} r^{1-2\gamma} \int_0^{2\pi} \lvert \widehat{f}(r,\theta) \rvert^2 \mathrm{d}\theta \mathrm{d}r,
	\end{align}
as shown in Appendix \ref{app:randomModels}. 
Finally, the SNR criterion \eqref{eq:SNR} for this background model becomes
	  {\begin{equation} \label{eq:SNR3}
		\SNR(f) = \frac{1}{\sigma^2} \frac{ \lvert \langle \widehat{T} ,\widehat{f} \rangle\rvert^2}{\lVert \widehat{g} \rVert_2^2}, \\
	\end{equation}	
	with $\widehat{g}(\bm{\omega}) = \widehat{f}  (\bm{\omega})  / \lVert \bm{\omega} \rVert^{\gamma}$.}
	We now present how to maximize this new criterion. \\

\vspace{-0.3cm}
	\begin{proposition} \label{prop:spectralshaping}
		A steerable filter $f$ with finite set of harmonics $H$ maximizes the SNR criterion \eqref{eq:SNR3} for the self-similarity order $\gamma \geq 0$  if and only if
	\begin{equation}\label{eq:fSNR2}
	 \widehat{f}(r,\theta) \,\propto\, r^{2\gamma} \sum_{n \in H} \widehat{T}_n (r) \mathrm{e}^{\mathrm{j} n \theta},
	\end{equation}
	where  $\widehat{T}_n$ are the Fourier radial profiles of $T$ given by \eqref{eq:integralTn}.  \\
	\end{proposition}
\vspace{-0.3cm}

Proposition \ref{prop:spectralshaping} is proved in Appendix \ref{ProofSpectralShaping}. The new optimal filter is simply the multiplication  in the Fourier domain of the optimal filter of Theorem \ref{theo:optimalfilter}  by $ r^{2\gamma}$, which corresponds to a \emph{spectral shaping} of \eqref{eq:fSNR}. 

	In practice, we do not necessarily know the self-similarity parameter \emph{a priori}.
	We can nevertheless estimate $\gamma$ from the background $S$ itself.
	The principle is as follows.
	For a test function $f$ and a scale $a>0$, we consider $a^{-1} f(\cdot /a)$ whose $L_2$-norm does not depend on $a$ and which allows  analyzing the background $S$ at scale $a$. 
	Then, the variance of $X_a = \langle S , a^{-1} f( \cdot / a) \rangle$ is known to be  proportional to $a^{2\gamma}$~\cite[Proposition 5.6]{Fageot2015wavelet}. 
	We therefore perform a multiscale analysis at various $a>0$ to estimate the parameter $\gamma$ from the theoretical linear relation between $\log \mathrm{Var}(X_a)$ and $\log a$, namely
	\begin{equation}
		\log \langle S , a^{-1} f( \cdot / a) \rangle = 2\gamma \log a + b,
	\end{equation}
	where $b \in \R$.
	This method was initially proposed in~\cite{Fageot2015wavelet} to analyse the statistics of natural images and shown to be robustly usable. 

	Finally, we remark that one may only have access to $I_0 = T + S$ in \eqref{eq:imageModel}, and not to the template $T$ itself, to design the optimal steerable filter. In that case, background substraction techniques can be used to recover $T$ from $I_0$. This aspect is not discussed further in the paper and we assume to have access to a good template representation.

\vspace{-0.3cm}	
		\subsection{Detection Procedure}
		\label{sec:detectionprocedure}
		The objective of the detection process is to reveal the positions and orientations of patterns corresponding to a template $T(\bm{x})$ in a (larger) image $I(\bm{x})$.
\review{We model the image $I$ as in \eqref{eq:detectionImageModelintro}.
Here, we assume that the template ${T}$, the variance, and the self-similarity of the ISS Gaussian field ${S}$ are known, while template locations $\boldsymbol{x}_j$ and orientations $\theta_j$ are unknown.}
In order to detect template locations $\bm{x}_j$ at the correct orientations $\theta_j$, we can efficiently compute the two following quantities using the steerability property~\eqref{eq:fsteerableRotation} of our detector $f$ as
\begin{align}
\label{eq:detectionArgmax}
I_{\text{ang}}(\bm{x}_0)&=\argmax_{\theta_0 \in [0, 2 \pi) }\langle I(\cdot-\bm{x}_0), f(\mathbf{R}_{\theta_0}\cdot) \rangle, \\
\label{eq:detectionMax}
I_{\text{amp}}(\bm{x}_0)&=\max_{\theta_0 \in [0, 2 \pi) }\langle I(\cdot-\bm{x}_0), f(\mathbf{R}_{\theta_0}\cdot) \rangle \\
&= \langle I(\cdot-\bm{x}_0), f(\mathbf{R}_{I_{\text{ang}}(\bm{x}_0)}\cdot) \rangle,  \nonumber
\end{align}
where $I_{\text{ang}}(\bm{x}_0)$ is the estimated orientation of $T$ at $\bm{x}_0\in\mathbb{R}^2$ and $I_{\text{amp}}(\bm{x}_0)$ is the amplitude of the maximum response of $f$ at $\bm{x}_0$.
It is worth noting that $I_{\text{amp}}(\bm{x}_j)$ will be maximized when $I_{\text{ang}}(\bm{x}_j)\approx\theta_j$.
				
\section{Optimal Steerable Filters: Discretization} \label{sec:OSFimplementation}

	We have now described how to deduce the optimal steerable filter associated to a template $T$ in a background $S$.
	The formulation of Section \ref{sec:theory} is in the continuous-domain,
	although images are in practice stored as discrete arrays in a computer. 
	The concrete design of detection algorithms therefore requires the discretization of the proposed theory.
	
	Practically, one should compute the optimal steerable detector $f$ in \eqref{eq:fSNR} from a discretized version of the template of interest $T$.
	Computing $f$ requires an angular averaging over the Fourier transform of the template in  \eqref{eq:integralTn}. 
	The approximation of \eqref{eq:integralTn} on the Cartesian grid is challenging because it involves evaluating integrals over the angular polar coordinate $\theta$ for all values of $r$, where much less samples are available when $r$ is small.
	
	\review{A strategy followed by~\cite{zhao2020fourier} consists of computing the Radon transform of $\widehat{f}$, from which one deduce the radial profils $\widehat{f}_n$~\cite[Section 2-D-1]{zhao2020fourier}. 
	This solution can be efficiently implemented and leads to excellent approximation performance. 
	However, the Radon transform must be computed for a large amount of angles in order to limit artefacts in the filtered back-projection reconstruction. }
	
		\review{Here, we propose an alternative discretization that relies on the expansion of the Fourier radial profiles in terms of radial B-splines, introduced in Section \ref{subsec:radialBsplines}. 
	We then combine the circular harmonics and radial B-splines in Section \ref{subsec:splineharmonic} and obtain the discretized optimal steerable filter in Theorem \ref{theo:Bsplineexpansion}. Finally, we summarize how to compute the discretized version of the optimal steerable filter in Section \ref{sec:integralInPractice}. 
	In comparison to the one involving Radon transforms, our solution relies on few B-spline coefficients. As such, it can be less memory demanding and may be useful for high-resolution images or large templates. The two methods are however implementing the same decomposition~\eqref{eq:ladecomposition} and are therefore similar in their goal. Both can be valuable, and the most convenient choice ultimately depends on the application context.}
	

\vspace{-0.3cm}	
	\subsection{B-Spline Expansion of the Radial Profiles}\label{subsec:radialBsplines}
	{Splines are widely used in signal processing because of their excellent approximation properties~\cite{deBoor1978practical}.
	Using spline interpolation, a function is approximated by polynomial pieces that are smoothly connected together.
	We consider the case of junctions between polynomials located on a grid, referred to as \emph{cardinal splines}~\cite{schoenberg1973cardinal,Unser2005think}. }
	The theory is traditionally developed for one-dimensional functions from $\R$ to $\R$, and has to be slightly modified in our context, since we deal with \emph{radial} functions.

	The B-spline of degree $0$ is given by $\beta_0 (x) = 1_{[- 1/2,1/2]} (x)$. The spline of degree $(M+1)$ is defined recursively as \mbox{$\beta_{M+1}(x) =  (\beta_M * \beta_0 ) (x)$}. In our experiments, we use the quadratic B-spline $\beta_2$, which is supported over $[-1.5,1.5]$ and is piecewise quadratic on the intervals $[k,k+1]$, $k \in \Z$. The closed form expression of  the quadratic B-spline can be found for instance in \cite{Unser1999splines}. Thereafter, we write $\beta_2 = \beta$ to simplify the notation.  \\

\vspace{-0.3cm}	
	\subsubsection{Radial B-splines}

A radial B-spline is a radial function $\widehat{f}(r)$ in $L_2(\R^2)$ of the form
\begin{equation} \label{eq:formradialspline}
	\widehat{f}(r) = \sum_{k \in \Z} \frac{c[k]}{r_0} \beta\left( \frac{r}{r_0} - k \right),
\end{equation}
with $r_0>0$   the discretization step and   $c[k]$ the spline coefficients of $\widehat{f}$. 
The function $\widehat{f}$ is defined for radius $r\geq 0$. One could  therefore restrict the sum in \eqref{eq:formradialspline} to integers $k$ such that the support of $\beta(\cdot / r_0 - k)$ intersects $\R^+$. For quadratic splines, this corresponds to $k \geq -2$. 
However, we prefer to keep the summation over all integers, allowing to consider discrete convolutions between sequences indexed by $k \in \Z$. 
In what follows, we approximate the Fourier radial profiles of the template $T$ using radial B-splines. 
\\

\vspace{-0.3cm}
	\subsubsection{Approximation with Radial B-Splines}

A  radial function $\widehat{f} \in L_2(\R^2)$ can be approximated by radial B-splines of the form \eqref{eq:formradialspline} at arbitrary precision by taking the discretization step $r_0 \rightarrow 0$.
This is well-known for classical B-splines \cite{Unser1999splines} and can be adapted to the case of radial B-splines.

The main difference between the usual B-splines expansion of 1D functions and the B-spline expansion of 2D radial functions is that one changes the scalar product. We recall that  the 2D scalar product between two radial functions $\widehat{f}$ and $\widehat{g}$ is given by
	\begin{align*}
		\langle \widehat{f} , \widehat{g} \rangle &=   
		 \int_0^\infty \int_0^{2\pi}  \widehat{f}(r) \overline{ \widehat{g}(r)} r  \mathrm{d} \theta  \mathrm{d} r  =  2 \pi \int_0^\infty  \widehat{f}(r) \overline{ \widehat{g}(r)} r \mathrm{d} r. 
	\end{align*}
	
	To facilitate   computations, we identify a radial function $\widehat{f} : \R^+ \mapsto \C$ to its symmetrization $\widehat{f} : \R \mapsto \C$ such that $\widehat{f}(-r) = \widehat{f}(r)$. In particular, the scalar product between two radial functions becomes
	$\langle \widehat{f} , \widehat{g} \rangle= \pi \int_{\R}  \widehat{f}(r) \overline{ \widehat{g}(r)} |r| \mathrm{d} r$.
	All the scalar products between radial functions have to be understood with this symmetrization procedure. 
	
	The expansion of a radial function in the quadratic spline basis requires   special attention since the family is not orthogonal. We can overcome this by using classical techniques  for B-splines that we adapt to the case of radial B-splines.
	We set $h[k] =  \langle \beta , \beta(\cdot - k) \rangle$ for each $k \in \Z$. The filter $h$ is nonzero only for $\lvert k \rvert\leq 2$ due to the support  of $\beta$. The fact that $h$ differs from the Kronecker   $\delta$ means precisely that the family of shifted B-spline is not orthonormal. 
	Since $h$ is compactly supported, there exists a unique discrete filter $h_{\mathrm{inv}}=(h_{\mathrm{inv}}[k])_{k\in \Z}$ such that $(h *h_{\mathrm{inv}}) [k] = ( h_{\mathrm{inv}} * h ) [k] = \delta[k]$~\cite{DeBoor1989,Fageot2017beyond}. 
	
	For any radial function $\widehat{f} \in L_2(\R^2)$, its projection to  the space of radial B-splines with discretization step $r_0$ is denoted by 
		\begin{equation} \label{eq:projectionBsplines}
			\mathrm{P}_{r_0} \{\widehat{f}\}(r) = \sum_{k \in \Z} \frac{c[k]}{r_0} \beta\left( \frac{r}{r_0} - k \right),
		\end{equation}
	
	\begin{proposition} \label{prop:radialexpansion}
		Let ${f} \in L_2(\R^2)$. 
		For $k\in \Z$, we set
		\begin{equation} \label{eq:thedk}
		d[k] = \frac{1}{2\pi} \left\langle \widehat{f} (r,\theta) , \frac{1}{r_0} \beta\left( \frac{r}{r_0} - k \right) \right\rangle.
		\end{equation}
		Then, the coefficients in the orthogonal projection \eqref{eq:projectionBsplines} of $\widehat{f}$ are computed as
		\begin{equation} \label{eq:cviad}
			c[k] = (h_{\mathrm{inv}} * d) [k].
		\end{equation}
	\end{proposition}	

Proposition \ref{prop:radialexpansion} is proved in Appendix \ref{app:Bsplineprop}. It gives the optimal approximation of a radial function in terms of radial B-splines for a given discretization step. Note that the discrete filter $h_{\mathrm{inv}}$  does not depend on the discretization step $r_0$. { Practically, it is computed once using basic signal processing tools, by adapting the principles exposed in~\cite{unser1991fast} to \emph{radial} B-splines. We first obtain the filter $h$ by computing the Gram matrix of the shifted B-splines, from which we deduce the Z-transform $H(z)$ of $h$, which is a polynomial. The Z-transform of $h_{\mathrm{inv}}$ is then given by $1/H(z)$, from which we can recover $h_{\mathrm{inv}}$ by identifying the poles of the rational function $1/H(z)$.}

\vspace{-0.3cm}	
	\subsection{Combining Radial B-splines and Circular Harmonics} \label{subsec:splineharmonic}
	
	We define the family of functions $\varphi_{n,k}$, $n,k \in \Z$, given in the Fourier domain by
		\begin{equation} \label{eq:phink}
			\widehat{\varphi}_{n,k} ( r , \theta ) = \frac{1}{r_0} \beta\left( \frac{r}{r_0} - k \right) \mathrm{e}^{\mathrm{j} n \theta}.
		\end{equation}
	For a fixed finite set of harmonics $H \subset \Z$ and discretization step $r_0 > 0$, one denotes by $\mathrm{P}_{r_0,H} \{\widehat{f}\}$ the orthogonal projection of the function $\widehat{f}$ onto the space generated by the $\widehat{\varphi}_{n,k}$ for $n \in H, k \in \Z$.
	
	We can combine Proposition \ref{theo:harmonicdecomposition} and Proposition \ref{prop:radialexpansion} to approximate any square-integrable function by steerable functions whose Fourier radial profiles are B-splines. \\
	
\vspace{-0.3cm}			
	\begin{theorem} \label{theo:Bsplineexpansion}
	Let $r_0 > 0$ and $H\subset \Z$. 
	For any function $T \in L_2(\R^2)$, the orthogonal projection of its Fourier transform on the $\widehat{\varphi}_{n,k}$, $n \in H, k \in \Z$
	is
		\begin{equation}
			\mathrm{P}_{r_0,H} \{\widehat{T}\}  (r , \theta ) = \sum_{n \in H} \sum_{k\in \Z} c_n[k] \widehat{\varphi}_{n,k} ( r, \theta),
		\end{equation}
		where $c_n[k] = (h_{\mathrm{inv}} * d_n) [k] = \sum_{\ell \in \Z} h_{\mathrm{inv}} [\ell] d_n [k -\ell]$ and
		\begin{align} \label{eq:coeffsdn}
			d_n[k] &= \frac{1}{2\pi}  \left\langle \widehat{T}  ,  \varphi_{n,k}  \right\rangle= \frac{1}{2\pi}  \left\langle \widehat{T}_n (r) , \frac{1}{r_0} \beta\left( \frac{r}{r_0} - k \right) \right\rangle.
		\end{align} 
		Moreover, when $r_0 \rightarrow 0$ and $H \rightarrow \Z$, the orthogonal projection converges to any $T$ for the $L_2$-norm. \\
	\end{theorem}
\vspace{-0.3cm}
	
	Theorem \ref{theo:Bsplineexpansion} is proved in Appendix \ref{app:Bsplineprop}. It allows to compute an approximation of any template $T$ from the B-spline coefficients $c_n[k]$ of the $n$th radial profile for each $n$. Moreover, this approximation can be as good as required by diminishing the step size $r_0$ and increasing the number of harmonics. These coefficients are obtained via the sequence $d_n$,  computing a simple convolution. Note that this operation is necessary because the family $\varphi_{n,k}$ is not orthogonal.
	In practice, one obtains the coefficients $d_n[k]$, and therefore   $c_n[k]$, by computing scalar products of the form \eqref{eq:coeffsdn}. 
	
	 Theorem \ref{theo:Bsplineexpansion} means that one can approximate the optimal steerable filter based on the integral \eqref{eq:coeffsdn}. This is a clear improvement since this 2D integral can be approximated from the knowledge of $T$ on a finite Cartesian grid. We develop this last point in the next section.

\vspace{-0.3cm}	
	\subsection{Computing the Discretized Optimal Steerable Detector}\label{sec:integralInPractice}
	
	In practice, we have access to the template $T$ in a finite pixel grid. The steps to compute the optimal steerable detector in Theorem \ref{theo:optimalfilter} are as follows. 
	\begin{itemize}
		\item Fix the set of harmonics $H$ and the discretization step $r_0$. 
		\item Compute the discrete Fourier transform $\widehat{T}$ of $T$ via   fast Fourier transform (FFT).
		\item Compute the coefficients $d_n[k]$ for $n\in H$ and $k \in K$, where $K$ is the set of integers such that $r_0 k$ remains in the range of the image. 			The scalar product \eqref{eq:coeffsdn} is expressed as an integral in Cartesian coordinates  as  
		\begin{equation} \label{eq:dncartesian}
			d_n[k] = \int_{\R^2} \widehat{T} (\omega_x,\omega_y) \widehat{\varphi}_{n,k} (\omega_x, \omega_y)  \mathrm{d} \omega_x \mathrm{d} \omega_y.
		\end{equation}
		This integral is approximated with its Riemann sum, from the knowledge of $\widehat{T}$ on the Cartesian grid. The expression of the basis functions $\widehat{\varphi}_{n,k}$ in Cartesian coordinates is
		\begin{small}
		\begin{equation}
			\widehat{\varphi}_{n,k}(\omega_x,\omega_y) = \frac{1}{r_0} \beta \left(\frac{\sqrt{\omega_x^2 + \omega_y^2}}{r_0} - k\right) \mathrm{e}^{\mathrm{j} n \arctan(\omega_y / \omega_x)}. 
		\end{equation}
		\end{small}
		\item For every $n \in H$ and $k \in K$, compute the $c_n[k]$ according to $c_n [k] = (h_{\mathrm{inv}} * d_n)[k]$. 
		\item Finally, the optimal spline-based steerable filter $f_{\mathrm{opt}}$ is given in the Fourier domain by
		\begin{equation} \label{eq:computefopt}
				\widehat{f}_{\mathrm{opt}} (r,\theta) = \sum_{k \in K} \sum_{n \in H} c_n[k] \frac{1}{r_0} \beta( r / r_0 - k) \mathrm{e}^{\mathrm{j} n \theta}.
		\end{equation}
	\end{itemize}
	
	We remark that the Riemann sum approximating the integral \eqref{eq:dncartesian} is obtained from a finite number of coefficients only. Indeed, it deals with the grid points lying in the area delineated by the radial function $\beta( r / r_0 - k)$. When the template is known only on a coarse grid, the quality of the estimation of $d_n[k]$ can therefore be insufficient. We address this issue by zero-padding the template $T$ in the spatial domain to increase the size of the image. This corresponds to a sinc interpolation in Fourier domain, increasing the number of points on which the integral \eqref{eq:dncartesian} is computed. 
	The relevance of our discretization method for the construction of the optimal steerable detector is investigated and illustrated in Section~\ref{sec:OSFpracticing}.
	
	When the background $S$ is adequately modeled as an ISS Gaussian field of self-similarity parameter $\gamma$ (see Section   \ref{subsec:spectralshaping}), the optimal detector is   characterized in Proposition \ref{prop:spectralshaping}. 
	It is simply obtained by multiplying  \eqref{eq:computefopt} by $r^{2\gamma}$. 
	We recall that the estimation of $\gamma$ can be performed efficiently on a single realization of the background $S$ using the method developed in \cite{Fageot2015wavelet}. 
	
\vspace{-0.3cm}
\section{Experimental Results} \label{sec:OSFpracticing}
In this section, we evaluate the performance and parameter sensitivity of the proposed optimal steerable filter design.
We first focus on template approximation, and follow with pattern detection and orientation estimation. \review{We finally demonstrate the usefulness of our method on real microscopy images.}

\vspace{-0.3cm}
	\subsection{Benchmarking Methodology} \label{subsec:comparison}
	
	\subsubsection{Competing Methods}	
		\review{To benchmark the performance of our method, we perform an experimental comparison with the two state-of-the-art approaches most closely related to it,~\cite{yang2013steerability} and~\cite{zhao2020fourier}. }
		
In~\cite{yang2013steerability}, the 2D Hermite kernel is defined as 
\begin{equation} \label{eq:phipqsigma}
\phi_{p,q,\sigma_{\text{Her}}}(\bm{x}) = \frac{1}{\sqrt{\sigma_{\text{Her}}}} H_p(x/\sigma_{\text{Her}}) H_q(y/\sigma_{\text{Her}}) \mathrm{e}^{ -  \lVert \bm{x} \rVert^2 /2\sigma_{\text{Her}}^2},
\end{equation}
where $\sigma_{\text{Her}}>0$ is a fixed scale parameter, $p,q \in \N$, and $H_p$ is the Hermite polynomial~\cite[Eq. (3)]{yang2013steerability}. The Hermite kernel  approximation of a template $T$ is obtained by projecting it onto the family $(\phi_{p,q,\sigma_{\text{Her}}})_{p+q \leq N_{\text{Her}}}$ for a fixed $N_{\text{Her}} \geq 0$, which specifies the Hermite-kernel-based filter used for the detection.
		This corresponds to using  $ (N_{\text{Her}} +1)(N_{\text{Her}} + 2)/2 $
		basis functions, to be compared with the $(N+1)$ elements used to generate our spline-based steerable detector, with $N$ the maximum harmonic. 
		We use equivalent numbers of basis functions when comparing the Hermite-based and spline-based methods. 

\review{In~\cite{zhao2020fourier}, steerable filters are constructed relying on the same representation as~\eqref{eq:ladecomposition}, but the radial profiles are obtained through a Radon transform. In the absence of spectral shaping, the only difference between this approach and ours therefore resides in the strategy adopted to compute the radial profiles, either via a Random transform or a radial B-spline approximation. For a fair comparison, we will hence use an identical number of basis functions $N$ for this method and ours in our experiments.}

{
\subsubsection{Dataset}\label{sec:dataset}
Detection tasks in which a pattern of interest can appear not only at any location in the image, but also at any orientation arise in biology, medicine, physics and chemistry~\cite{Kawamoto2013,Beck2004,Guichard2013,Mader2012,frangi1998multiscale,DFA2017,depeursinge2018rotation,PTB2009}.
However, to the best of our knowledge, no publicly available dataset allows benchmarking algorithms on such a task.
We therefore created artificial datasets in order to carry out controlled experiments and evaluate our method against the state-of-the-art. We used templates $T$ and backgrounds $S$ to generate two microscopy-like image collections featuring templates appearing at random locations $\bm{x}_j$ and random orientations $\theta_j$, subject to various background intensity levels.

The images $I$ are generated following the mathematical model \eqref{eq:detectionImageModelintro}.
The templates include an elongated curved structure resembling the digit three,  $T_{\text{three}}$, and a double helix (DH) point spread function $T_{\text{DH}}$~\cite{PTB2009}. They are depicted in Fig.~\ref{fig:approx}.
The dimensions of the templates are $200\times 200$ pixels.
$T_{\text{three}}$ was designed to allow the evaluation of detection performance on non-polar-separable templates.
$T_{\text{DH}}$ is a point spread function used in a microscopy technique in which the PSF orientation encodes the depth of a fluorescence signal~\cite{PTB2009}.
$T_{\text{three}}$ and $T_{\text{DH}}$ are blended into backgrounds built from histopathological images\footnote{$4422 \times 2934$ image of \emph{plasmodium falciparum}, courtesy of Dr. M. D. Hicklin, public domain.}
 and realizations of ISS Gaussian fields ($\gamma=1.2$), respectively (Fig.~\ref{fig:templateBlending}).
\begin{figure}[t!]
\centering
\includegraphics[scale=0.16]{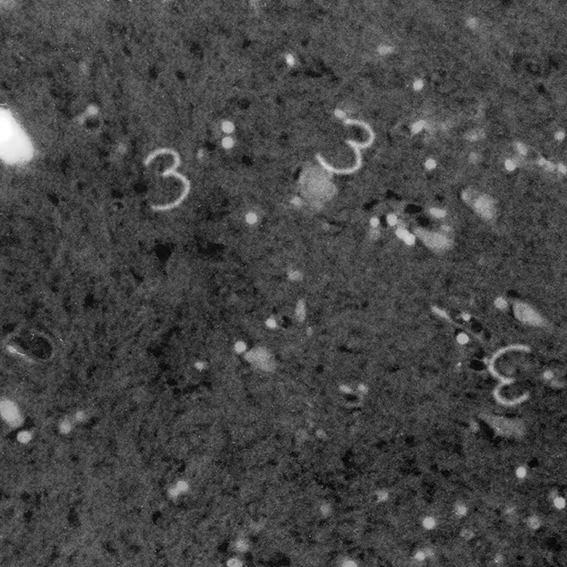}
\includegraphics[scale=0.16]{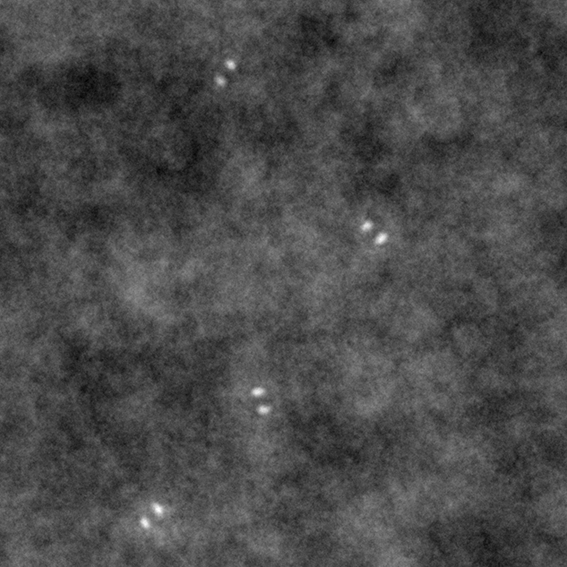}
\caption{Controlled experimental dataset to evaluate detection performance.
Left: $T_{\text{three}}$ in histopathological background ($1200 \times 1200$). Right: $T_{\text{DH}}$ in an ISS Gaussian field ($1200 \times 1200$, $\gamma=1.2$).}
\label{fig:templateBlending}
\end{figure}
}

\vspace{-0.3cm}
\subsection{Template Approximation}\label{sec:templateApprox}
	We first evaluate the approximation error of our spline-based steerable filter when compared with the template $T$. This scenario corresponds to a background with flat power spectrum (Theorem~\ref{theo:optimalfilter}).
	The optimal steerable detector is computed following the steps described in Section~\ref{sec:integralInPractice}.
	Both qualitative and quantitative results are showed in Fig.~\ref{fig:approx} for $T_{\text{three}}$ (a) and $T_{\text{DH}}$ (b).
	The approximation performance is measured in terms of the root mean squared error (RMSE).
	
	The parameter $r_0$ in~(\ref{eq:computefopt}) is determined for each template independently as a trade-off between achieving a fine-grained resolution for the interpolation of $r$ and having a sufficient number of coefficients in the Riemann sum approximating the integral~(\ref{eq:dncartesian}).
	In all experiments, $r_0$ values of $0.033$ and $0.041$ were used for $T_{\text{three}}$ and $T_{\text{DH}}$, respectively.
	{The parameter $r_0$ is inversely proportional to the number of radial splines used for the reconstruction (range of the parameter $k$ in \eqref{eq:phink}). It is the result of a trade-off: a high $r_0$ value leads to poor approximation performances for the radial splines, but too small values are problematic for the Riemann sum estimation of \eqref{eq:dncartesian}.
	
	\review{As we here focus on investigating the benefit of shaping the radial profile of the filters to the template, we only compare our method with the one based on Hermite kernels. 
	Since the template approximation obtained by~\cite{zhao2020fourier} is constructed following the same representation (see Proposition~\ref{theo:harmonicdecomposition}), 
	we focus on comparing with the resulting detection performance of this method in the following section.} We report the RMSE of Hermite kernel-based approximation for different number of basis functions in Fig.~\ref{fig:approx}. The parameter $\sigma_{\text{Her}}> 0$ in \eqref{eq:phipqsigma} controls the scale at which one expects templates to be detected. In our experiments, we selected the optimal $\sigma_{\text{Her}}$ value relying on grid search. Detection quality is  highly sensitive to the choice of this parameter. Although the role and sensitivity of $\sigma_{\text{Her}}$ is similar to that of our method's $r_0$, the parameter $\sigma_{\text{Her}}$ is part of the continuous-domain theory of Hermite kernels, while $r_0$ originates from the spline-based discretization procedure in our case. 
	
	In~Fig.~\ref{fig:approx}~(c), the RMSE between the original template and its approximation is plotted as a function of the number of basis functions used to design the corresponding steerable filters (see Section \ref{subsec:comparison}).
	This reveals that our spline-based method has excellent approximation properties compared to the Hermite-kernel-based one. This is not totally surprising since optimal radial profiles are designed to minimize the SNR. Additional information on the approximation properties of both frameworks can be found in the Supplementary Material.}
	
	Fig.~\ref{fig:approx}  also shows that, although using more harmonics consistently reduces the approximation error, only a small number of harmonics (\emph{e.g.}, $N\approx 6$) is required to accurately model the template in our method.
	Such a small amount of harmonics yields a low-pass approximation of the templates in terms of circular frequencies, which is striking for $T_{\text{three}}$ (see Fig.~\ref{fig:approx}~(a)).
	For $T_{\text{DH}}$, the RMSE decreases with $N$ in a less regular way, and the second harmonic is of paramount importance to model the two characteristic diametrically opposed blobs of the DH.
	The RMSE decays much faster for $T_{\text{DH}}$, which is circularly smoother than $T_{\text{three}}$.
	Fig.~\ref{fig:approx} (c) also reveals that a Fourier-based approximation is better-suited for regular patterns such as the double helix.

	\begin{figure}[t!]
	\begin{minipage}{\columnwidth}
	\centering
	\includegraphics[scale=0.19]{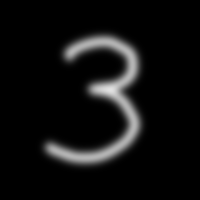}\hspace{4pt}
	\includegraphics[scale=0.19]{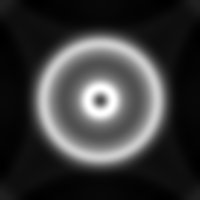}
	\includegraphics[scale=0.19]{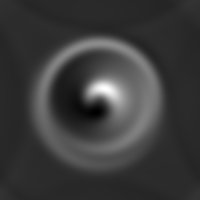}
	\includegraphics[scale=0.19]{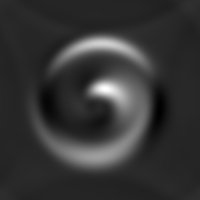}
	\includegraphics[scale=0.19]{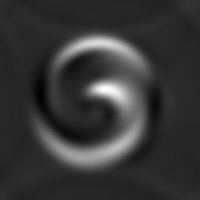}
	\includegraphics[scale=0.19]{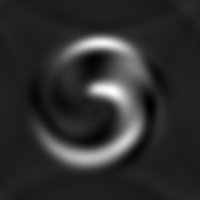}
	\end{minipage}
	\begin{minipage}{\columnwidth}
	\centering
	\vspace{4pt}
	\subfloat[Three: original $T_{\text{three}}$ and approximations (Spline) $f$ with $N=0,1,\dots,9$.]{
	\hspace{39pt}
	\includegraphics[scale=0.19]{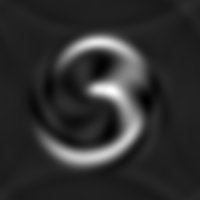}
	\includegraphics[scale=0.19]{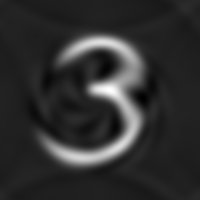}
	\includegraphics[scale=0.19]{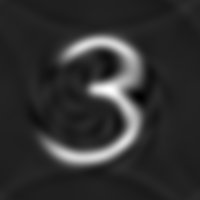}
	\includegraphics[scale=0.19]{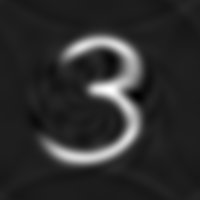}
	\includegraphics[scale=0.19]{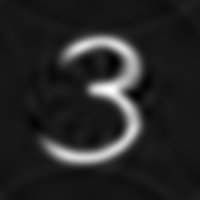}
	}
	\end{minipage}
	\begin{minipage}{\columnwidth}
	\vspace{6pt}
	\centering
	\includegraphics[scale=0.19]{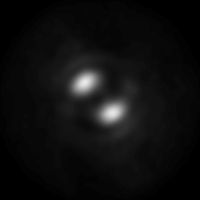}\hspace{4pt}
	\includegraphics[scale=0.19]{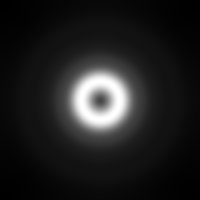}
	\includegraphics[scale=0.19]{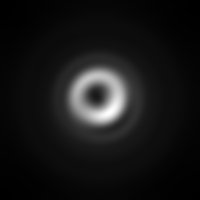}
	\includegraphics[scale=0.19]{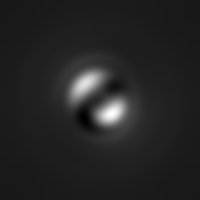}
	\includegraphics[scale=0.19]{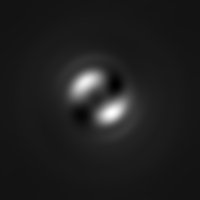}
	\includegraphics[scale=0.19]{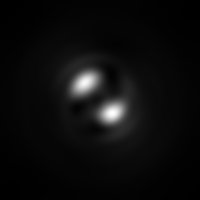}
	\end{minipage}
	\begin{minipage}{\columnwidth}
	\centering
	\vspace{4pt}
	\subfloat[DH: original $T_{\text{DH}}$ and approximations (Spline) $f$ with $N=0,1,\dots,9$.]{
	\hspace{39pt}
	\includegraphics[scale=0.19]{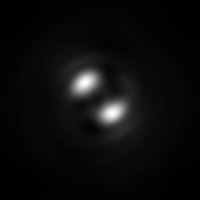}
	\includegraphics[scale=0.19]{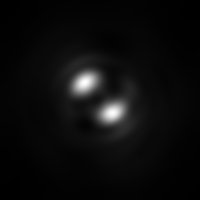}
	\includegraphics[scale=0.19]{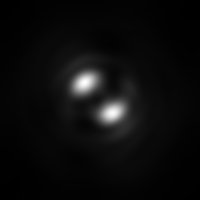}
	\includegraphics[scale=0.19]{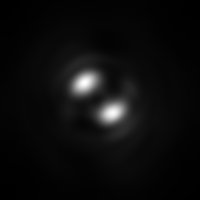}
	\includegraphics[scale=0.19]{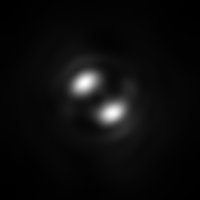}
	}
	\end{minipage}
	\vspace{2pt}
	\subfloat[{Evolution of template approximation error (RMSE) with respect to the number of basis functions used.}]{
	\includegraphics[scale=0.52]{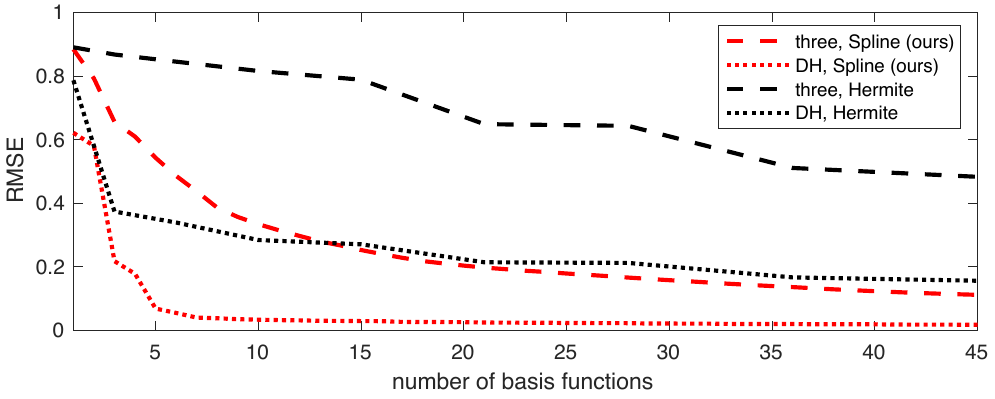}
	}
	\caption{{Influence of the number of basis functions for template approximation.}}
	\label{fig:approx}
	\end{figure}

\vspace{-0.3cm}
\subsection{Estimation of Template Position and Orientation} \label{subsec:detect}
 \subsubsection{Experimental Setup and Evaluation Metrics}
 {
	For the following experiments, we rely on the synthetic dataset described in Section~\ref{sec:dataset}.
A template's orientation is estimated according to \eqref{eq:detectionArgmax}, while the amplitude of the maximum response of the filter $f$ at $\bm{x}_0$, defined in \eqref{eq:detectionMax}, is used as detection score. A  detection score and angle is thus retrieved at each pixel in the images $I_\mathrm{amp}$ and $I_\mathrm{ang}$, respectively.
We use a strict detection criterion in which only a detection matching the pixel at the center of the template is considered as true positive.
It is worth noting that, since the ratio of true positives to true negatives is small (\emph{e.g.}, 1/500,000), receiver operating characteristic (ROC) analysis is inadequate as it mostly focuses on the ability of the system to find the true positions of the template. Precision-recall (PR) analysis, on the other hand, better captures the system's tendency to generate false detections. We therefore rely on PR curves to evaluate detection performance.
In all experiments, we provide areas under the PR curve (AUC) with and without spectral shaping (SS), referred to as ``Spline + SS (ours)'' and ``Spline (ours)'', respectively.
The self-similarity parameter $\gamma$ used for the spectral shaping is estimated using the method described in the last paragraph of Section~\ref{sec:integralInPractice}, based on 10 realizations of $S(\bm{x})$.}

Unless mentioned otherwise, we fix a level of background intensity of $\sigma=1$ and test \review{all methods} for $M=30$ angle values uniformly distributed in $[0,2\pi)$.
The self-similarity parameters $\gamma$ were estimated to be $1.35$ and $1.21$ for the histopathological images and the ISS Gaussian fields, respectively.
A total of $N=20$ and $N=8$ harmonics were used  to approximate the templates $T_{\text{three}}$ and $T_{\text{DH}}$, respectively.

\subsubsection{Robustness to Background Intensity}
We first investigate the robustness of detection performance in various levels of background intensity for $T_{\text{three}}$, in Fig.~\ref{fig:detectionNoiseLevel_Three}.
{The use of spectral shaping grants excellent robustness to noise with a PR AUC well above all other approaches for all levels of background intensity. 
An angular error of less than 10 degrees is observed for a background intensity level of $\sigma=5$, while the template can hardly be seen under such a SNR.}
This highlights the adequacy of the isotropic and self-similar model for the histopathological background.
The evolution of the PR AUC and the corresponding cropped response maps suggest that most false positives occur in the vicinity of the true template position.
The resulting false positive could then be excluded by local non-maximum suppression.

\review{We compare results obtained with Hermite-kernel-based steerable detectors~\cite{yang2013steerability}, Fourier-Argand steerable detectors constructed with the Radon transform~\cite{zhao2020fourier}, and steerable detectors obtained with our approach. Hermite-kernel-based steerable detectors achieve significantly worse performance than non-spectrally shaped detectors (constructed with the Radon transform or with radial B-splines), suggesting that our construction, equivalent to the Fourier-Argand from~\cite{zhao2020fourier}, is better suited for the detection task considered in Fig.~\ref{fig:detectionNoiseLevel_Three}. This tendency is less pronounced for $T_{\text{DH}}$ (Fig.~\ref{fig:detectionNoiseLevel_DH}), which can be attributed to the fact that this template is initially better approximated by Hermite kernels (as seen in Fig.~\ref{fig:approx}~(c)). For detection, the steerable detector of ~\cite{zhao2020fourier} are only slightly outperformed by ours in the low-noise regime when no spectral shaping is considered. This competing method however results in larger angular errors, especially for $T_\mathrm{DH}$. This may be attributed to the fact that the radial B-spline expansion slightly regularizes the template approximation procedure. In both experiments, the positive impact of spectral shaping is striking. }

When performing a detection task relying on filtering, a classical issue comes from the direct current (DC) component of the input image. If both the filter and background are not locally zero mean, the filter will indeed strongly respond in regions where the background is bright, leading to false positives. 
A classical procedure for dealing with such situations is to suppress the background of the input image in small overlapping neighbourhoods, making it effectively of zero local mean (LBS, discussed in Section \ref{sec:intro}). 
Spectral shaping offers a more elegant way of addressing this problem by suppressing low frequencies through windowing in the Fourier domain with $r^{2\gamma}$. 
For the sake of completeness, we repeated the experiments of Fig.~\ref{fig:detectionNoiseLevel_DH} on LBS images and detail them in the Supplementary Material. The results show that, as expected, LBS improves the performance of both Hermite-kernel-based and spline-based steerable filters  (ours) without spectral shaping. Our proposed spectral shaping approach however improves detection results far beyond what can be achieved with LBS.

\begin{figure}[t!]
\centering
\includegraphics[trim = 120 315 120 325, clip, scale=0.63]{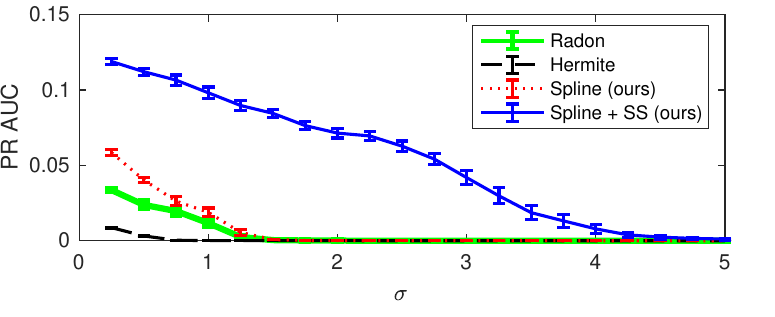}
\begin{minipage}{\columnwidth}
\hspace{51pt}
\includegraphics[trim = 292 368 170 99, clip, scale=0.32]{3smo_200_data1_ori_tapisInv_2_a=1_rea1_ori}
\hspace{-1pt}
\includegraphics[trim = 292 368 170 99, clip, scale=0.32]{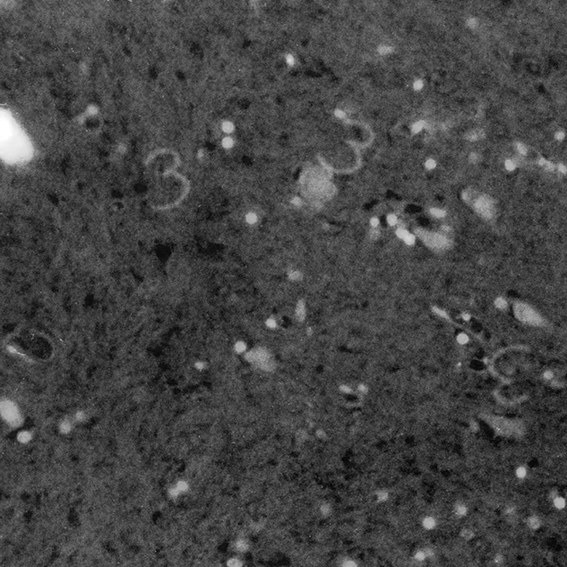}
\hspace{-1pt}
\includegraphics[trim = 292 368 170 99, clip, scale=0.32]{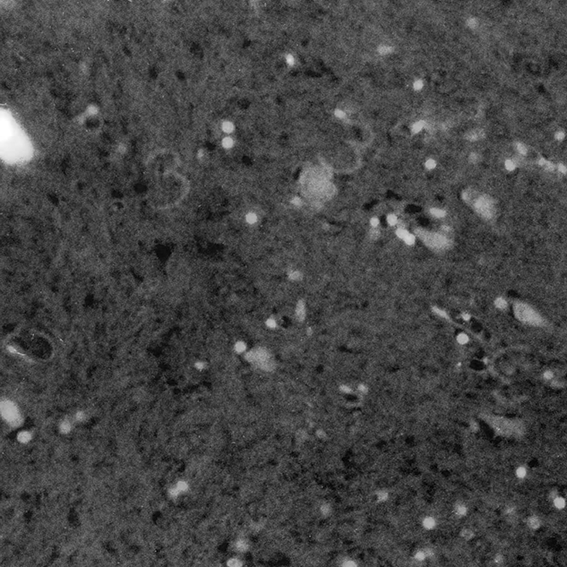}
\hspace{-1pt}
\includegraphics[trim = 292 368 170 99, clip, scale=0.32]{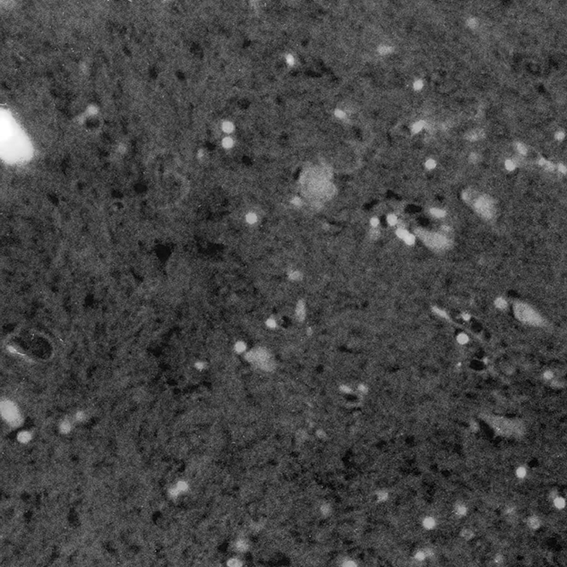}
\hspace{-1pt}
\includegraphics[trim = 292 368 170 99, clip, scale=0.32]{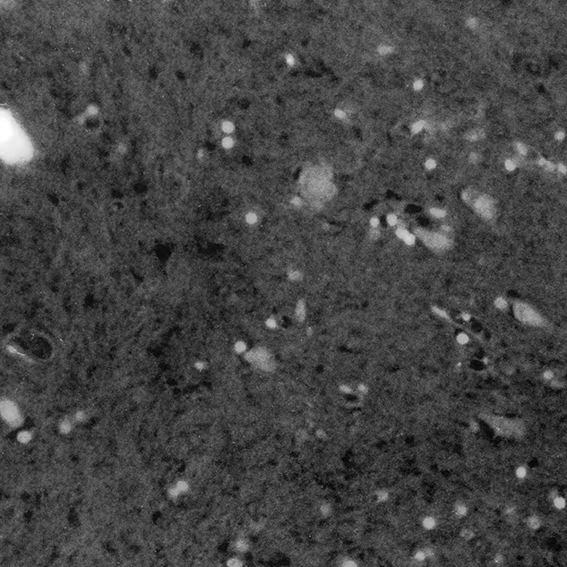}
\put(-227,12){\small $I^{\text{loc}}(\bm{x})$}
\end{minipage}
\begin{minipage}{\columnwidth}
\vspace{4pt}
\hspace{51pt}
\includegraphics[trim = 292 368 170 99, clip, scale=0.32]{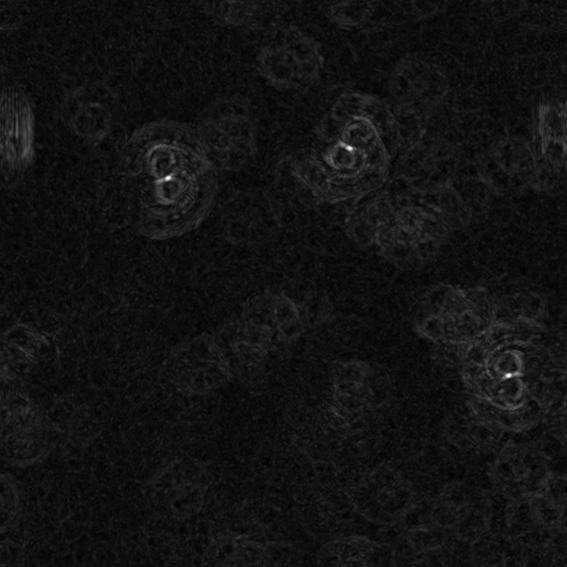}
\hspace{-1pt}
\includegraphics[trim = 292 368 170 99, clip, scale=0.32]{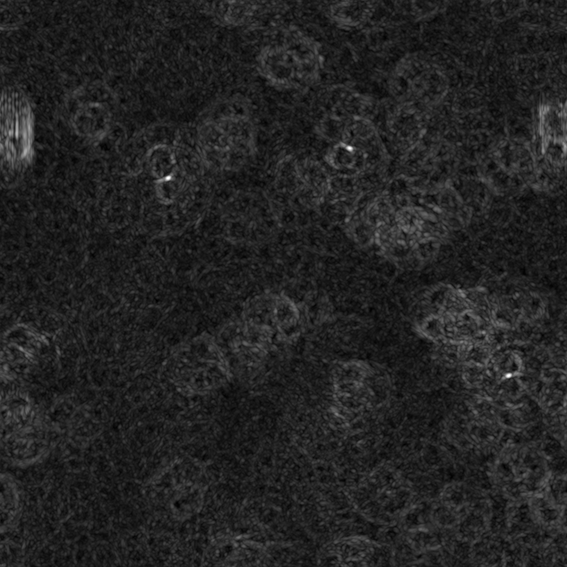}
\hspace{-1pt}
\includegraphics[trim = 292 368 170 99, clip, scale=0.32]{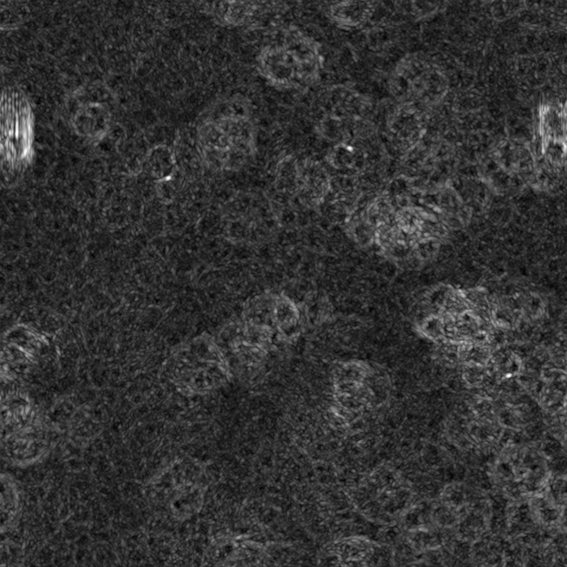}
\hspace{-1pt}
\includegraphics[trim = 292 368 170 99, clip, scale=0.32]{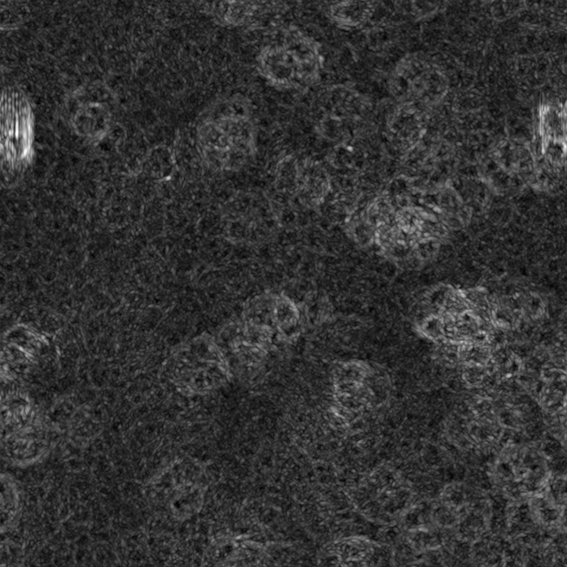}
\hspace{-1pt}
\includegraphics[trim = 292 368 170 99, clip, scale=0.32]{3smo_200_data1_ori_tapisInv_2_a=4_rea1_ori_amp_white}
\put(-229,16){\small $I^{\text{loc}}_{\text{amp}}(\bm{x})$}
\put(-232,7){\scriptsize(Spline+SS)}
\end{minipage}
\begin{minipage}{\columnwidth}
\vspace{4pt}
\hspace{51pt}
\includegraphics[trim = 292 368 170 99, clip, scale=0.32]{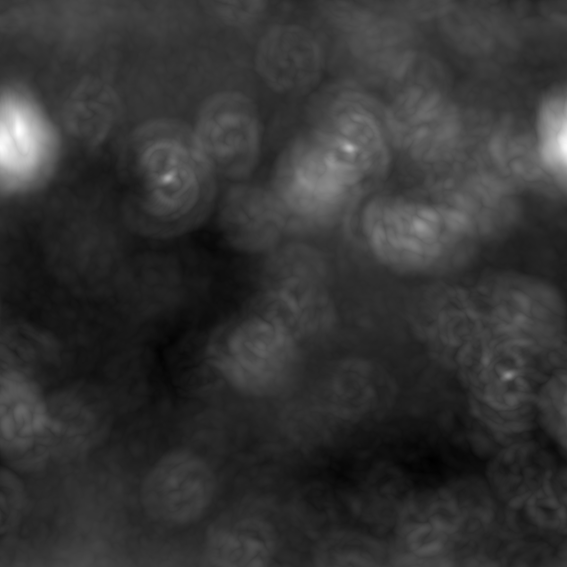}
\hspace{-1pt}
\includegraphics[trim = 292 368 170 99, clip, scale=0.32]{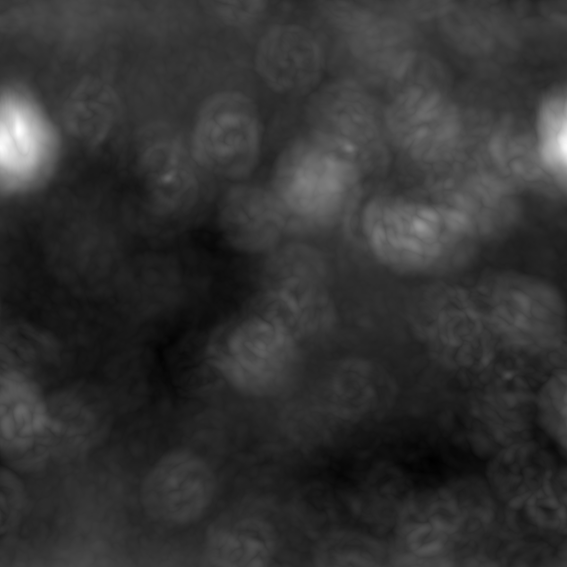}
\hspace{-1pt}
\includegraphics[trim = 292 368 170 99, clip, scale=0.32]{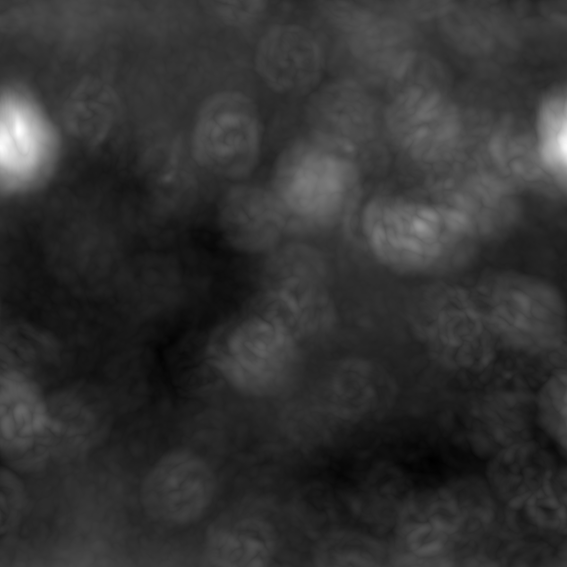}
\hspace{-1pt}
\includegraphics[trim = 292 368 170 99, clip, scale=0.32]{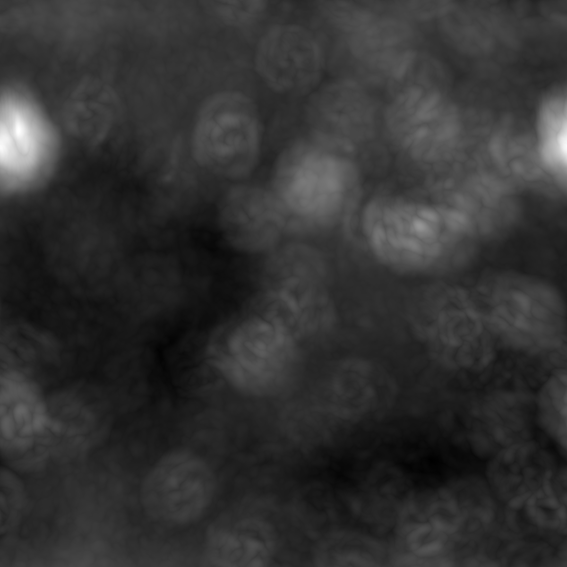}
\hspace{-1pt}
\includegraphics[trim = 292 368 170 99, clip, scale=0.32]{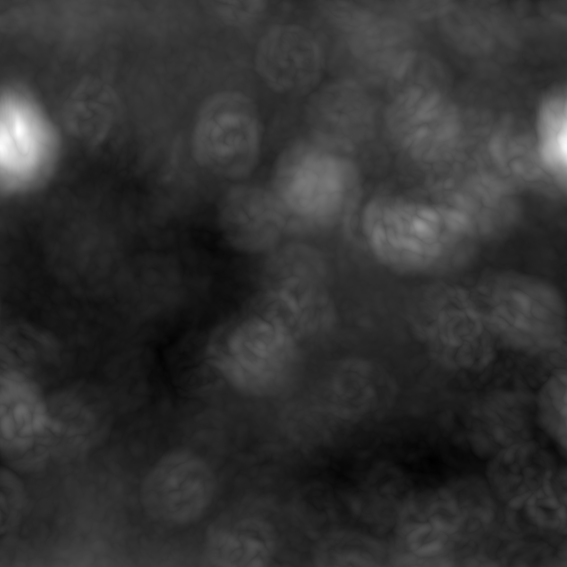}
\put(-229,16){\small $I^{\text{loc}}_{\text{amp}}(\bm{x})$}
\put(-228,7){\scriptsize{(Spline)}}
\end{minipage}
\begin{minipage}{\columnwidth}
\vspace{4pt}
\hspace{51pt}
\includegraphics[trim = 292 368 170 99, clip, scale=0.32]{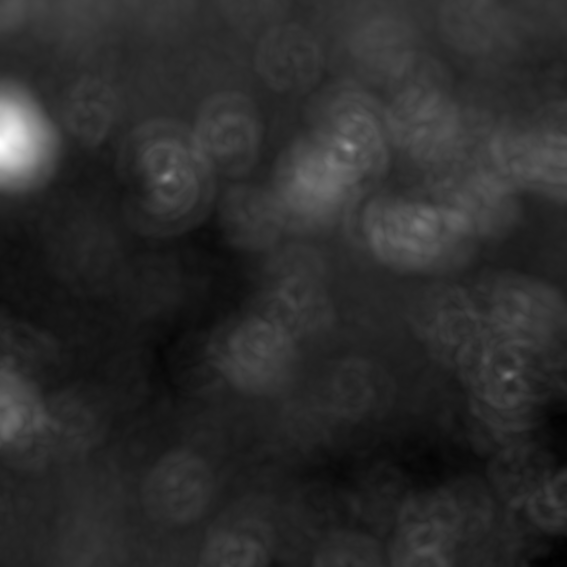}
\hspace{-1pt}
\includegraphics[trim = 292 368 170 99, clip, scale=0.32]{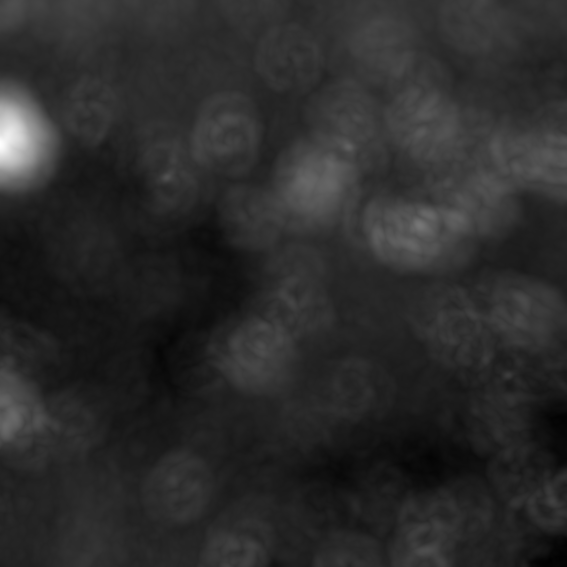}
\hspace{-1pt}
\includegraphics[trim = 292 368 170 99, clip, scale=0.32]{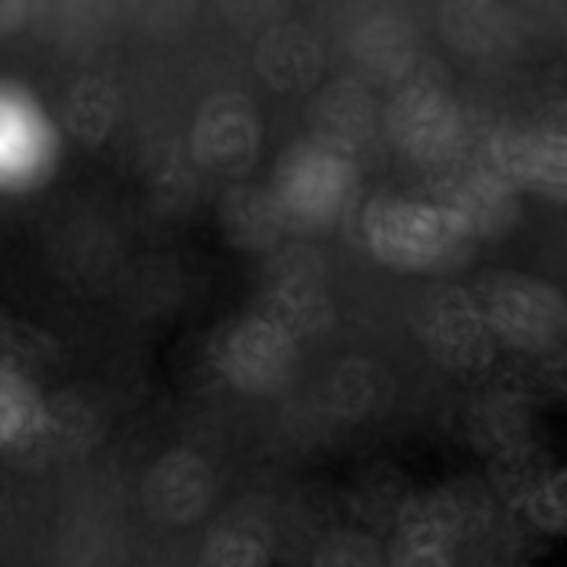}
\hspace{-1pt}
\includegraphics[trim = 292 368 170 99, clip, scale=0.32]{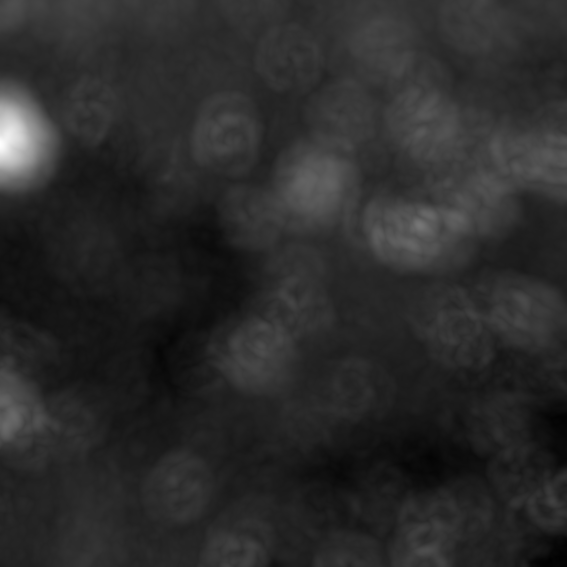}
\hspace{-1pt}
\includegraphics[trim = 292 368 170 99, clip, scale=0.32]{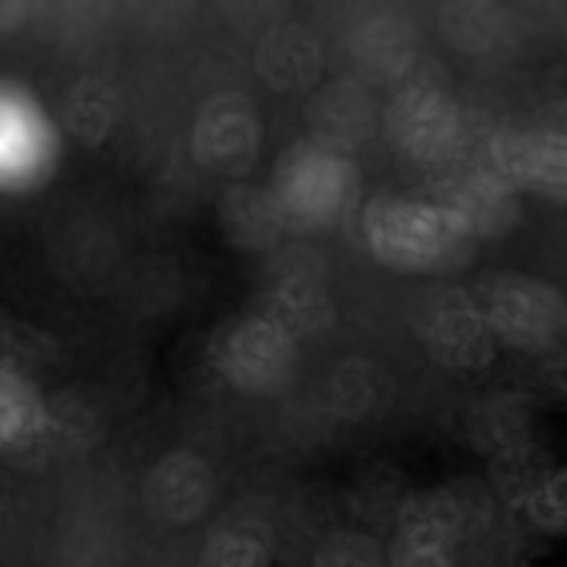}
\put(-229,16){\small $I^{\text{loc}}_{\text{amp}}(\bm{x})$}
\put(-228,7){\scriptsize{(Radon)}}
\end{minipage}
\begin{minipage}{\columnwidth}
\vspace{4pt}
\hspace{51pt}
\includegraphics[trim = 292 368 170 99, clip, scale=0.0455]{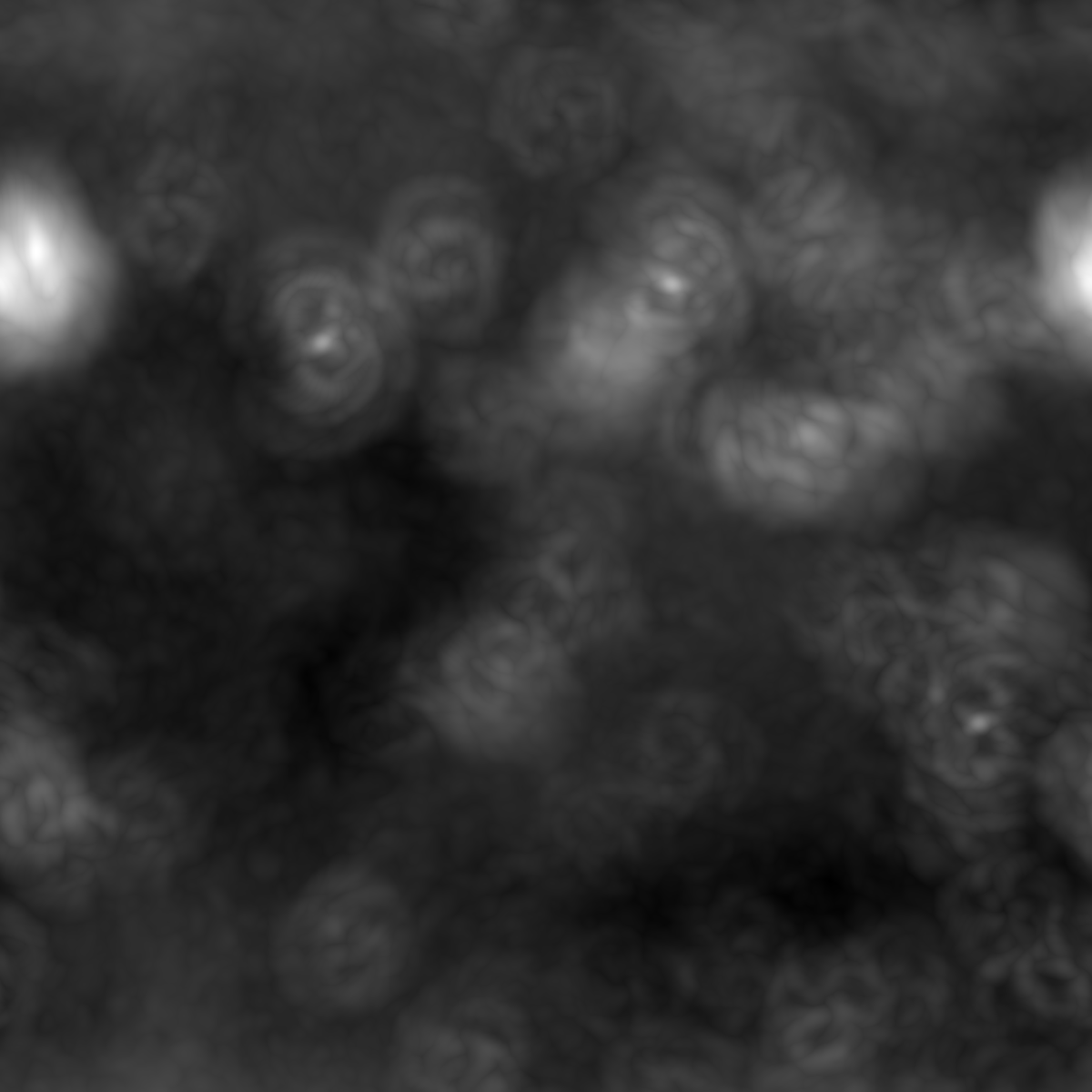}
\hspace{-1pt}
\includegraphics[trim = 292 368 170 99, clip, scale=0.0455]{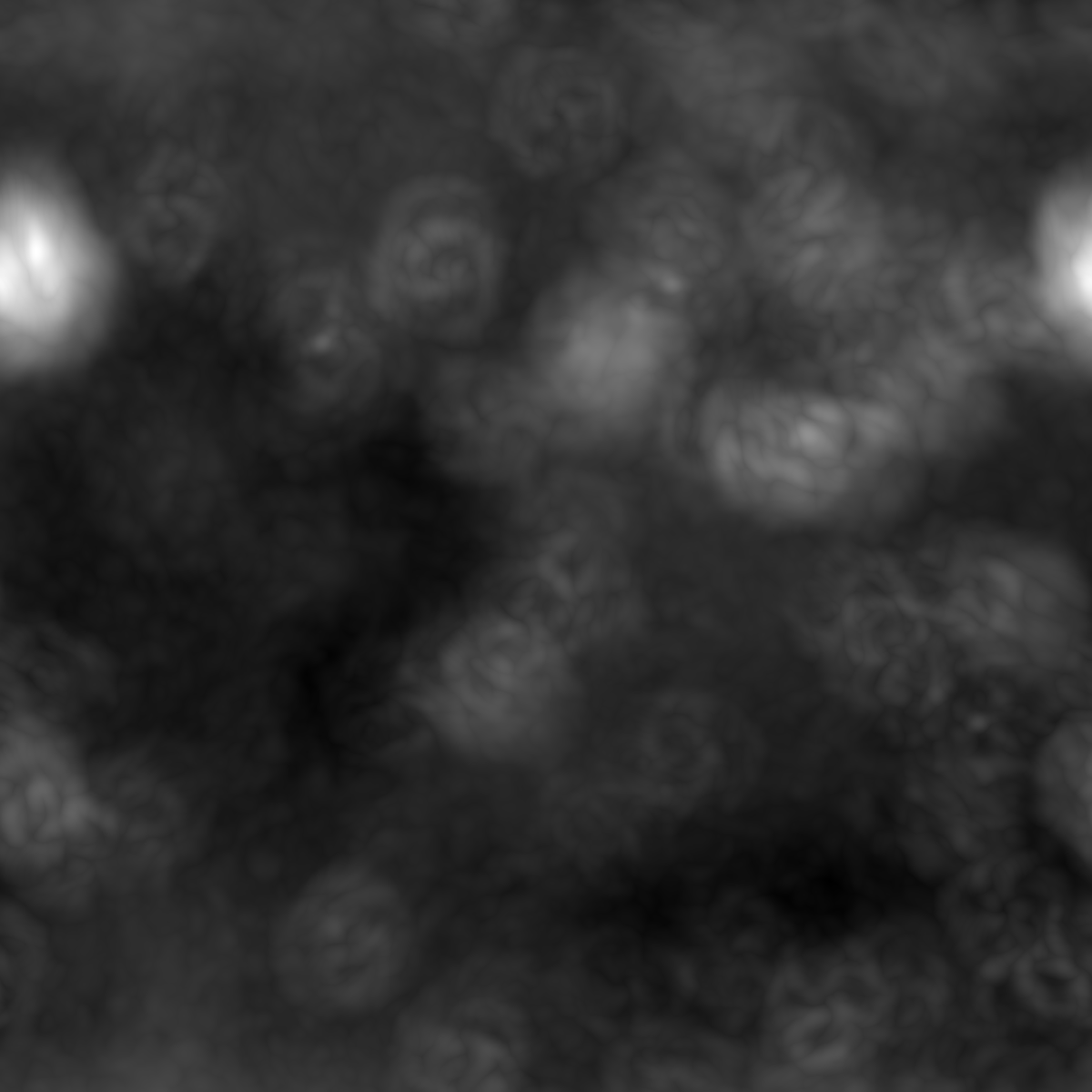}
\hspace{-1pt}
\includegraphics[trim = 292 368 170 99, clip, scale=0.0455]{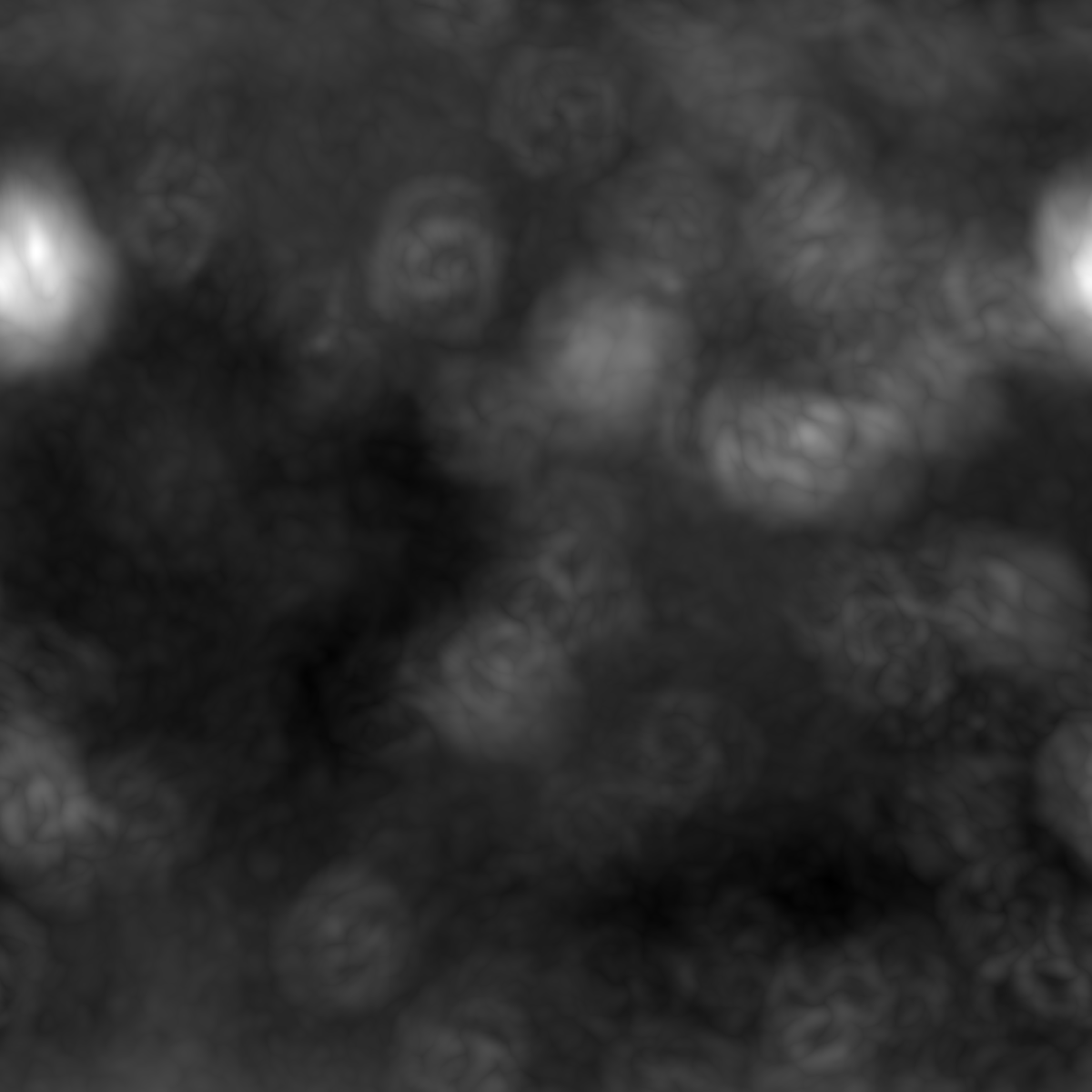}
\hspace{-1pt}
\includegraphics[trim = 292 368 170 99, clip, scale=0.0455]{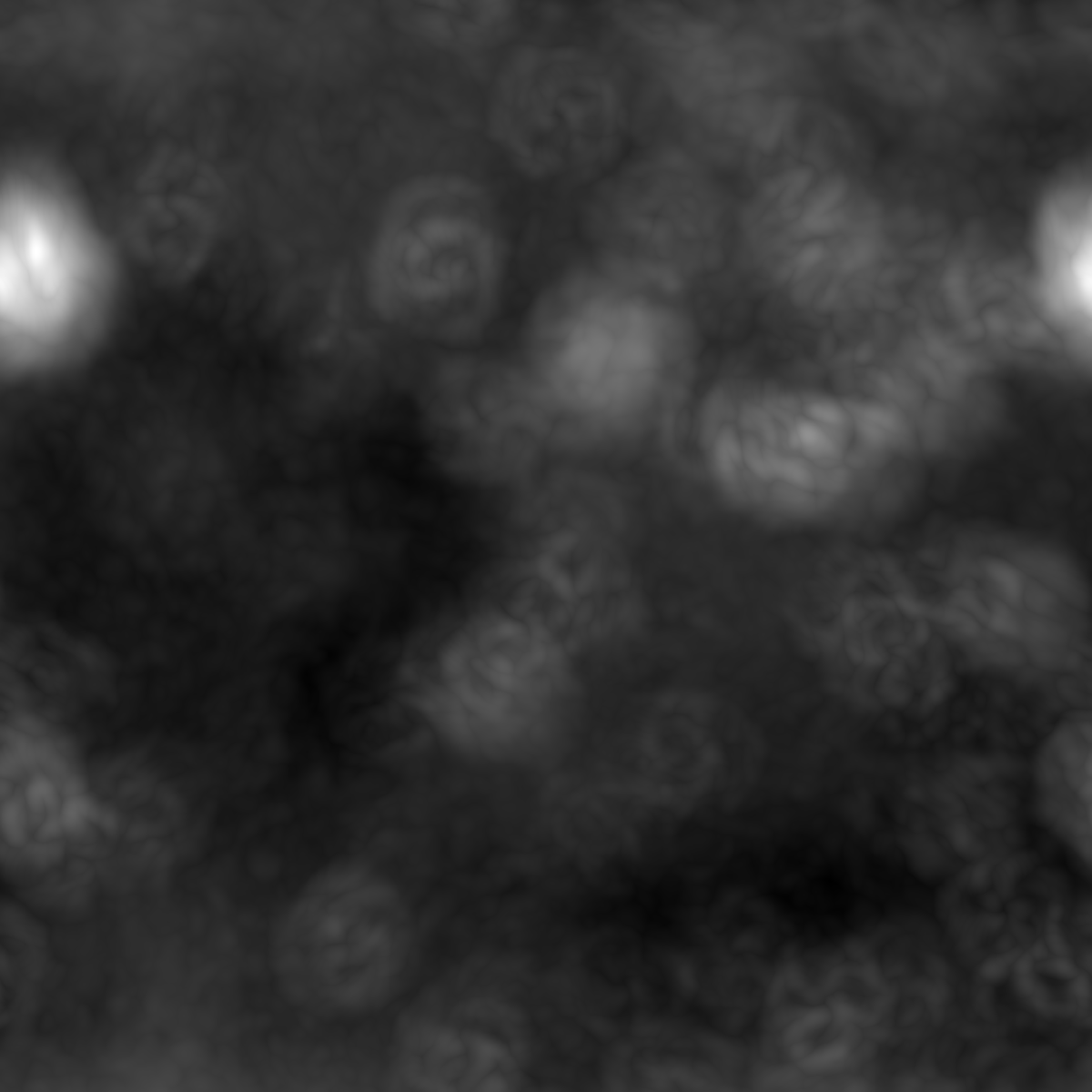}
\hspace{-1pt}
\includegraphics[trim = 292 368 170 99, clip, scale=0.0455]{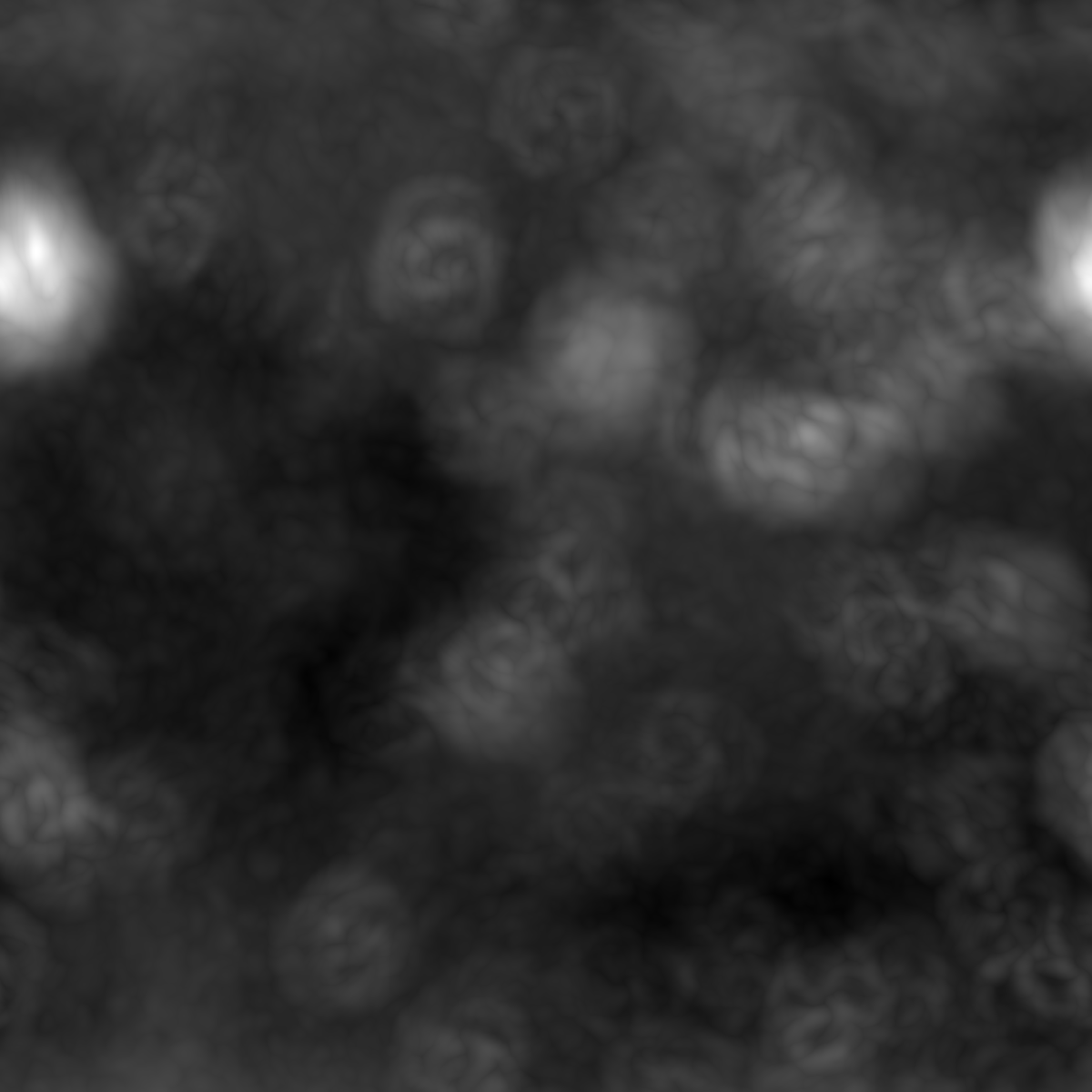}
\put(-229,16){\small $I^{\text{loc}}_{\text{amp}}(\bm{x})$}
\put(-230,7){\scriptsize(Hermite)}
\end{minipage}
\begin{minipage}{\columnwidth}
\hspace{51pt}
\vspace{8pt}
\end{minipage}
\includegraphics[trim = 120 320 120 325, clip, scale=0.63]{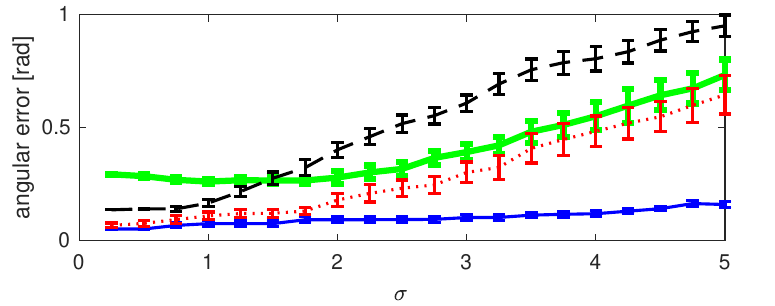}
\caption{Robustness of detection of $T_{\text{three}}$ for various levels of background intensity (histopathological background).
Cropped thumbnails of the image $I^{\text{loc}}$, and amplitude response $I^{\text{loc}}_{\text{amp}}(\bm{x})$ are shown to illustrate and compare the spatial distribution of the detection scores around a true positive for the corresponding level of background intensity $\sigma=1,\dots,5$.
{\color{black}The performance of equivalent steerable detectors built either from Hermite kernels ($N_{\text{Her}}=14,\,\, \sigma_{\text{Her}}=15$) 
or from the Fourier-Argand representation using the Radon transform (Radon) are reported for comparison. The angular errors are computed for $M=30$ angle values.}}
\label{fig:detectionNoiseLevel_Three}
\end{figure}
\begin{figure}[t!]
\centering
\includegraphics[trim = 120 332 120 325, clip, scale=0.63]{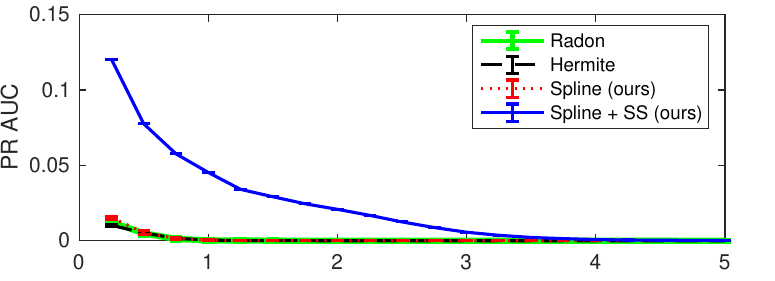}
\includegraphics[trim = 120 320 120 325, clip, scale=0.63]{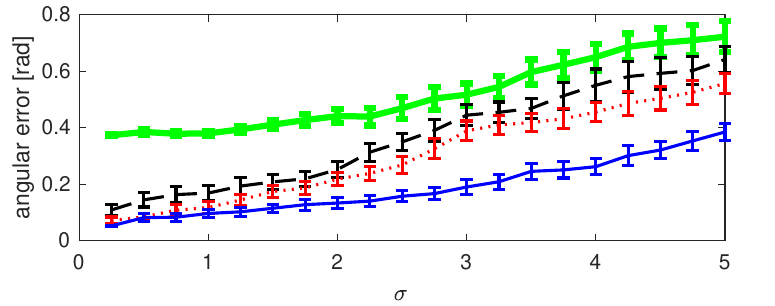}
\caption{Robustness of detection of $T_{\text{DH}}$ for various levels of background intensity (ISS Gaussian fields).
{\color{black}The performance of equivalent steerable detectors built either from Hermite kernels ($N_{\text{Her}}=8,\,\, \sigma_{\text{Her}}=15$) or from the Fourier-Argand representation using the Radon transform (Radon) are reported for comparison.  The angular errors are computed for $M=30$ angle values.}}
\label{fig:detectionNoiseLevel_DH}
\end{figure}

\subsubsection{Robustness to Self-similarity Parameter $\gamma$}
The robustness of the estimation of the self-similarity parameter $\gamma$ and its influence on the detection performance is studied in Table~\ref{tab:gammaEstimate} and Fig.~\ref{fig:gammaEstimateDetection}.
The estimated values $\tilde{\gamma}$ are found to be robust to the presence of templates for both types of background, which suggests that the  self-similarity  parameter can be directly estimated from $I(\bm{x})$ when the template density is relatively low.
A value of $\tilde{\gamma}=1.2$ corresponds to the ground truth for ISS Gaussian fields.
The optimal values for the detection of $T_{\text{three}}$ and $T_{\text{DH}}$ were found to be $\gamma_{\text{three, opt}} = 1.2$ and $\gamma_{\text{DH, opt}} = 1.4$, respectively, which are close to $\tilde{\gamma}$.
Even if the correspondence between $\tilde{\gamma}$ and $\gamma_{\text{opt}}$ is remarkable, a precise estimation of $\gamma$ is not found to be critical as the PR AUC plateaus around $\gamma_{\text{opt}}$.
It is worth noting  that, when detecting $T_{\text{DH}}$ in the ISS Gaussian field, the assumption of white background (\emph{i.e.}, no spectral shaping) leads to poor detection performance.
This is consistent with the findings in Fig.~\ref{fig:detectionNoiseLevel_DH}:
the spectral shaping operation becomes essential with this type of background (ISS Gaussian fields).
\begin{table}
\caption{Estimated self-similarity parameters $\tilde{\gamma}$ are compared when based on the pure background $S(\bm{x})$ versus $I(\bm{x})$ containing the templates~(\ref{eq:detectionImageModelintro}).
The values yielding optimal detection performance $\gamma_{\text{opt}}$ are compared.}
\label{tab:gammaEstimate}
\begin{center}
\begin{tabular}{c|c|c|c|}
$S(\bm{x})$ & $\tilde{\gamma}$ from $S(\bm{x})$ & $\tilde{\gamma}$ from $I(\bm{x})$ & $\gamma_{\text{opt}}$ \\
\hline
\begin{minipage}{60pt}
\vspace{3pt}
\centering
ISS
Gaussian fields ($\gamma=1.2$)\\
\vspace{3pt}
\end{minipage}
 & 1.2 & 1.21 & 1.2 \\
\hline
\begin{minipage}{60pt}
\vspace{3pt}
\centering
histopathological\\
images\\
\vspace{3pt}
\end{minipage}
 & 1.31 & 1.35 & 1.4 \\
\hline
\end{tabular}
\end{center}
\end{table}
\begin{figure}[t!]
\centering
\includegraphics[trim = 120 320 120 325, clip, scale=0.63]{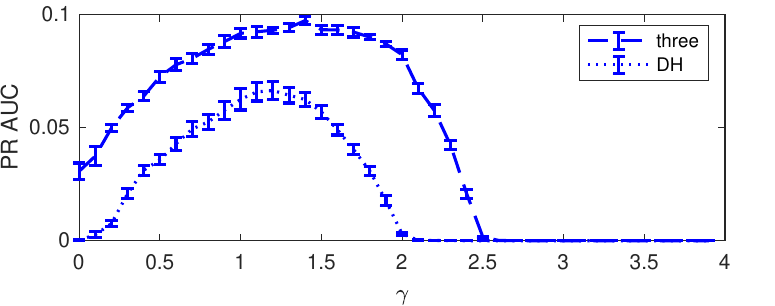}
\caption{Evolution of the PR AUC as a function of the self-similarity parameter $\gamma$ for $T_{\text{three}}$ and the histopathological background as well as for $T_{\text{DH}}$ and the ISS Gaussian field.
$\gamma_{\text{opt}}$ values of 1.4 and 1.2 corresponds to the optimal PR AUC for $T_{\text{three}}$ and $T_{\text{DH}}$, respectively.
In both cases, a precise estimation is not critical as the AUC plateaus around $\gamma_{\text{opt}}$.}
\label{fig:gammaEstimateDetection}
\end{figure}

\subsubsection{Impact of Total Number of Harmonics}
We investigated the importance of the number of harmonics $N$ on template approximation in Section~\ref{sec:templateApprox}.
The impact of the latter on detection performance is shown in Fig.~\ref{fig:detectionNbHarmDH} for $T_{\text{DH}}$.
The observed AUC and angular errors are consistent with our previous observations, where the importance of harmonic $N=2$ is highlighted to capture the two blobs of $T_{\text{DH}}$.
The performance is stable for $N\geq 4$, and using more harmonics does not significantly improve the detection.
Similar observations  are made on the influence of $N$ for detecting $T_{\text{three}}$.
\begin{figure}[t!]
\centering
\includegraphics[trim = 120 332 120 325, clip, scale=0.63]{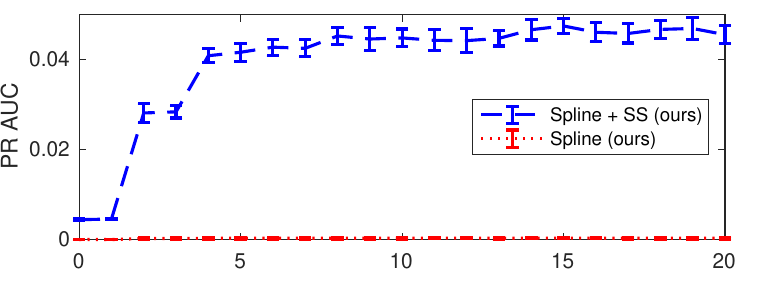}
\includegraphics[trim = 120 320 120 325, clip, scale=0.63]{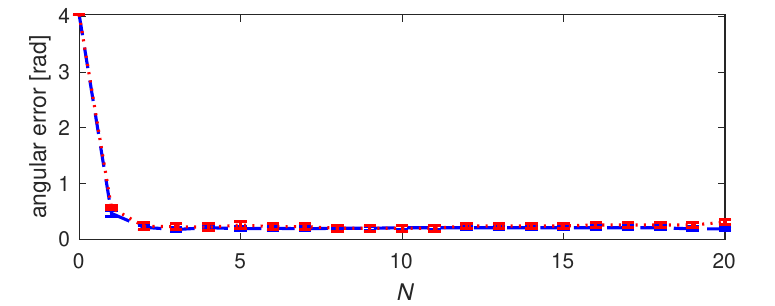}
\caption{Impact of the number of harmonics $N$ on the detection of $T_{\text{DH}}$. The findings are consistent with the study of template approximation error in Fig.~\ref{fig:approx}, where the second harmonic is capturing the two distinctive blobs of the DH.}
\label{fig:detectionNbHarmDH}
\end{figure}

\subsubsection{Impact of Total Number of Detection Angles}
We report the influence of the number $M$ of tested angles in $[0,2\pi)$ for \eqref{eq:detectionArgmax} and~\eqref{eq:detectionMax} in Fig.~\ref{fig:detectionTestedAngles_three} for $T_{\text{three}}$. On the same plot, we also indicate the \emph{baseline} angular error corresponding to the sampling step $\pi/M$.
Whereas performance consistently improves with $M$, relatively coarse angular discretization (\emph{e.g.}, $M=12$) yields near to optimal detection rates. 
Once again, very similar trends were observed on the influence of $M$ for detecting $T_{\text{DH}}$ in ISS Gaussian fields.
\begin{figure}[t!]
\centering
\includegraphics[trim = 120 332 120 325, clip, scale=0.63]{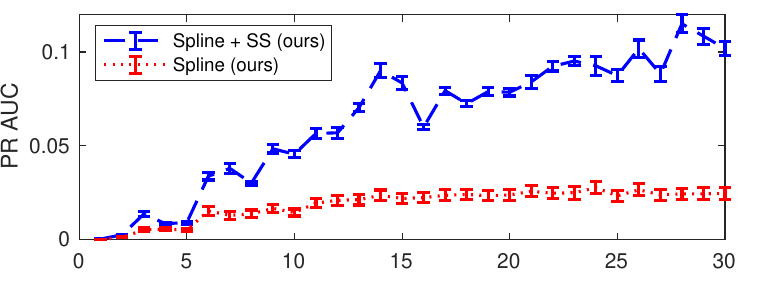}
\includegraphics[trim = 120 320 120 325, clip, scale=0.63]{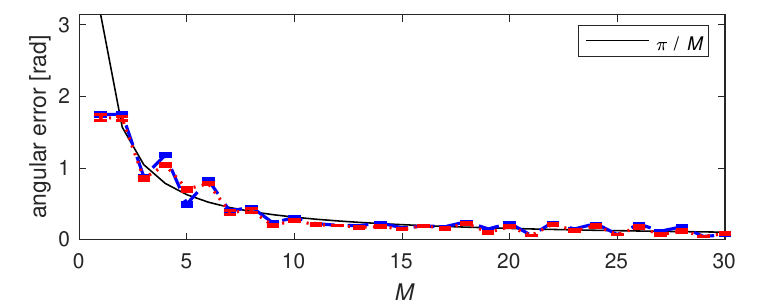}
\caption{Number $M$ of angles tested in $[0,2\pi)$ for the construction of $I_{\text{ang}}$ and $I_{\text{amp}}$ when detecting $T_{\text{three}}$.
The baseline angular error $\pi/M$ is reported.}
\label{fig:detectionTestedAngles_three}
\end{figure}

\subsection{\review{Application to Real Data Including Template Variations}}
\review{To showcase the use of our optimal steerable filter design in real-life conditions, we experiment on publicly available bioimage data from the Cell Image Library. We consider transmission electron microscopy images of the internal structure of motile cilia and flagella. These biological structures are responsible for the propelling movement observed for instance in sperm cells, which is crucial for assessing male fertility. They have a remarkably conserved organization composed of microtubule doublets arranged into a highly symmetrical bundle.}

\review{In Figures~\ref{fig:bio1} and~\ref{fig:bio2}, we illustrate how we can build a steerable filter that detects microtubule doublets from a single template in two different datasets. As no ground truth annotation is available for these data, we however have to limit ourselves to a qualitative evaluation of performance. In spite of differences in appearance due to image noise and morphological variations of the microtubules, most doublets are efficiently detected and their orientation appears to be correctly estimated, as seen from the orientation of the radius depicted inside each detection circle.
Our approach could therefore be used for instance to quantify changes of microtubule doublet arrangements in mutants exhibiting motility dysfunctions.}

\begin{figure}[t!]
\centering
\includegraphics[width=\linewidth]{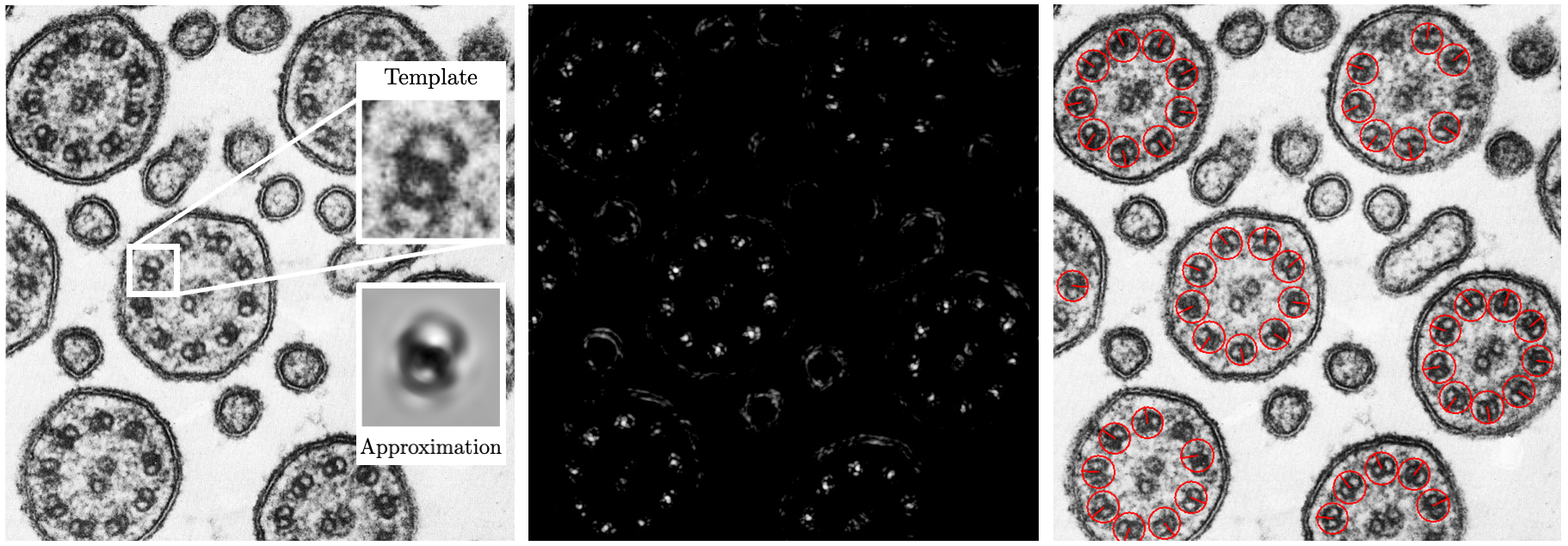}
\caption{\review{Cross-section of rat tracheal epithelial cells (Cell Image Library, CIL:11623). Left: an image template is approximated (Spline+SS) with $N=6$ harmonics. A self-similarity parameter of $\gamma=0.5$ is used for the spectral shaping. Center: amplitude response of the optimal steerable filter. Right: detection (circles) and orientation (radius inside circle) results.}}
\label{fig:bio1}
\end{figure}

\begin{figure}[t!]
\centering
\includegraphics[width=\linewidth]{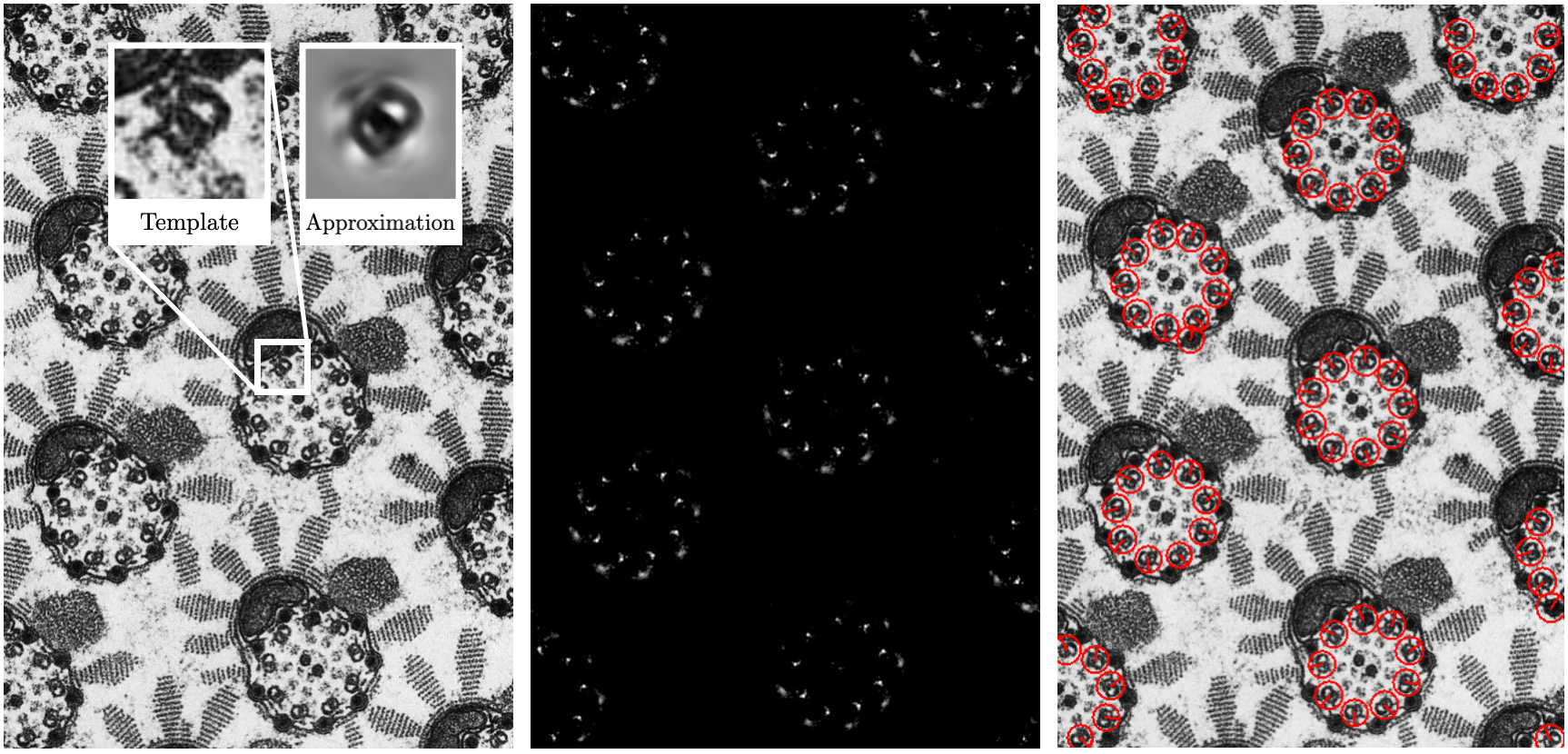}
\caption{\review{Cross-section of \emph{Lepidoptera} sperm cells (Cell Image Library, CIL:35962). Left: an image template is approximated (Spline+SS) with $N=6$ harmonics. A self-similarity parameter of $\gamma=0.25$ is used for the spectral shaping. Center: amplitude response of the optimal steerable filter. Right: detection (circles) and orientation (radius inside circle) results.}}
\label{fig:bio2}
\end{figure}

\section{Discussion and conclusion} \label{sec:discussandconclude}

The main goal of this work is to provide a complete pipeline for the detection of specific patterns at unknown positions and orientations in an image. 
The key ingredients of our approach are (a) a continuous-domain formulation of the detection problem based on steerable filters and the optimization of the SNR criterion (Section~\ref{sec:theory}),
 (b) a  radial  spline-based discretization scheme (Section~\ref{sec:OSFimplementation}),
and (c) a novel spectral shaping approachß to extract statistical information from the background  noise (the self-similarity parameter $\gamma$) and specify an appropriate image model. 
We have experimentally studied our detection procedure's ability to adequately estimate the position and orientation of a pattern of interest (Section \ref{sec:OSFpracticing}). 
We here briefly recap the main contributions of our work.

\begin{itemize}

	\item We approximated a template with steerable filters, which provides a low-pass approximation retaining only  small angular frequencies.
	In practice, we have seen that only few angular frequencies are required for  good template approximation (Fig. \ref{fig:approx}) and pattern detection (Fig. \ref{fig:detectionNbHarmDH}). 

	
	\item Our method relies on the assumption that the patterns of interest in the image are adequately modeled by a single template $T$ provided by the user, as in \eqref{eq:detectionImageModelintro}.
		Under this assumption, we are able to precisely detect the occurrences of this template (Section~\ref{subsec:detect}) with a strong robustness to noise (Fig.~\ref{fig:detectionNoiseLevel_Three} and~\ref{fig:detectionNoiseLevel_DH}).
		
	\item \review{When compared to state-of-the-art steerable-filter-based approaches (Hermite kernels~\cite{yang2013steerability} and Fourier-Argand detectors~\cite{zhao2020fourier}), our method performs consistently better than its alternatives without spectral shaping.}	
Spectral shaping (Section \ref{subsec:spectralshaping}) tremendously improves detection performance (Fig.~\ref{fig:gammaEstimateDetection}). This demonstrates the relevance of considering background models with non-flat spectrum. The proposed isotropic and self-similar model only requires the estimation of one parameter, $\gamma$, for which we provide a theoretically justified procedure. 
	{ This strategy is consistent with the more classical one consisting of using LBS as preprocessing step, but extends it beyond the sole removal of the zero-frequency.}
	
	\item We are not only able to recover the positions of the patterns, but also their \emph{a priori} unknown orientation. This information is simply extracted from the angular map $I_{\text{ang}}$ in \eqref{eq:detectionArgmax}. To refine angular accuracy, one should consider enough test angles. Higher values of $M$ do not significantly improve detection performance but obviously affect the angular error (Fig. \ref{fig:detectionTestedAngles_three}). This additional angular information could be exploited in segmentation~\cite{Uhlmann2016hermite} or to extract directional features of the object of interest, \emph{e.g.}, in bioimage analysis (Fig.~\ref{fig:bio1} and~\ref{fig:bio2}).
	
\end{itemize}

Limitations of the current approach includes the modeling of one single template per detector. 
However, this potential weakness is compensated by the ability of the model to approximate any template with high accuracy.
A collection of detectors can thus be obtained by generating a dedicated detector for each distinct template class.
Detection robustness with respect to pattern deformations was not explored as it lies out of the scope of the problem we consider in this work.
Nevertheless, we believe that using steerable models with a small number of harmonics results in regularized (i.e., low-pass) detectors capturing the global layout of the template with enhanced generalization abilities.
\vspace{-0.2cm}
\appendix
\vspace{-0.1cm}

\subsection{Polar Decomposition of $L_2$-functions}
\label{ProofDecomposition}
	We provide  here the proof of Proposition \ref{theo:harmonicdecomposition}.
We consider a function $f \in L_2(\R^2)$ {\color{black} and its Fourier transform $\widehat{f} \in L_2(\R^2)$}. For $r \geq 0$ fixed, the function $\theta \mapsto \widehat{f} (r,\theta)$ is in $L_2([0,2 \pi))$ and can therefore be decomposed in Fourier series. We denote by $\widehat{f}_n(r)$ the complex Fourier coefficients, such that, for all $\theta\in [0,2\pi)$, 
\begin{equation}
	{\color{black}\widehat{f} (r,\theta)} = \sum_{n\in \Z}\widehat{f}_n(r) \ee^{\uj n \theta}. 
\end{equation}
Using Parseval's relation and the orthogonality of the system $(\widehat{f}_n(r) \ee^{\uj n \theta})_{n\in \Z}$, we have that
\begin{align*}
	\lVert f \rVert^2_2 &=  \frac{1}{2\pi} \lVert \widehat{f} \rVert_2^2 
	 = \frac{1}{2\pi} \int_0^\infty \int_0^{2\pi} \lvert \widehat{f}(r,\theta)\rvert^2 r {\rm d} \theta {\rm d} r  \\
			& = \frac{1}{2\pi} \int_0^{\infty} \left( \int_0^{2\pi} \sum_{n,m\in \Z} \widehat{f}_n(r) \overline{\widehat{f}
			_m(r)} \ee^{\uj (n-m) \theta} {\rm d} \theta \right) r{\rm d} r \\
			& =  \int_0^\infty \sum_{n \in \Z} \lvert \widehat{f}_n(r) \rvert^2 r{\rm d} r = \sum_{n\in\Z} \lVert \widehat{f}_n \rVert_2^2.
\end{align*}
This proves that $f$ is square-integrable if and only if $ \sum_{n \in \Z}  \lVert \widehat{f}_n \rVert_2^2 < \infty$. Finally, we remark that, for every $r\geq 0$, 
\begin{align*}
\frac{1}{2\pi} \int_0^{2\pi} \widehat{f} (r, \theta) \ee^{- \uj n \theta} {\rm d} \theta &= \sum_{m\in \Z}  \widehat{f}_m(r) \frac{1}{2\pi} \int_0^{2\pi}   \ee^{\uj (m-n) \theta} {\rm d} \theta \\
&= \widehat{f}_n(r),
\end{align*}
which proves \eqref{eq:fn} together with the fact that the decomposition is unique.

\vspace{-0.3cm}
\subsection{Optimal Steerable Filter: Without Spectral Shaping}
\label{ProofOptiFilter}
This section is dedicated to the proof of Theorem  \ref{theo:optimalfilter}.
Using Proposition \ref{theo:harmonicdecomposition}, we decompose the template in Fourier domain as $\widehat{T}(r,\theta) = \sum_{n\in \Z} \widehat{T}_n (r) \mathrm{e}^{\mathrm{j} n \theta}$. 
We set
\begin{equation}
\mathrm{P}_H \{\widehat{T} \}(r,\theta) = \sum_{n\in H} \widehat{T}_n (r) \mathrm{e}^{\mathrm{j} n \theta},
\end{equation}
where $H$ is the finite set of harmonics.
The operator $\mathrm{P}_H$ corresponds to the orthogonal projection onto the space of steerable filters with a set of harmonics $H$.
Then, the orthogonality properties of the circular harmonics $\mathrm{e}^{\mathrm{j} n \theta}$ easily implies that $\langle \widehat{f}, \widehat{T} \rangle = \langle \widehat{f} , \mathrm{P}_H \{\widehat{T}\} \rangle$ when $\widehat{f}$ is steerable with set of harmonics $H$. The Cauchy-Schwarz inequality now implies that $\lvert \langle \widehat{f}, \mathrm{P}_H \{\widehat{T} \} \rangle\rvert \leq \lVert \widehat{f} \rVert_2 \lVert \mathrm{P}_H \{ \widehat{T}\} \rVert_2$. 
Putting things together, one therefore has
\begin{align}
\mathrm{SNR}(f) &= \frac{\vert \langle T , f \rangle\rvert^2 }{\lVert f \rVert_2^2} = \frac{\vert \langle \widehat{T} ,\widehat{f} \rangle\rvert^2}{\lVert \widehat{f} \rVert_2^2}= \frac{\vert \langle \mathrm{P}_H \{\widehat{T}\} ,\widehat{f} \rangle\rvert^2}{\lVert \widehat{f} \rVert_2^2} \nonumber \\
&\leq \lVert \mathrm{P}_H\{\widehat{T}\} \rVert_2^2. \label{eq:boundSNRforproof}
\end{align}
Moreover, the upper bound in \eqref{eq:boundSNRforproof}, that does not depend on $f$, is reached if and only if $\widehat{f}$ is proportional to $\mathrm{P}_H\{\widehat{T}\}$ (equality case in the Cauchy-Schwarz inequality), as  expected. Note that the above reasoning is valid because the function $\mathrm{P}_H\{\widehat{T}\}$ is itself steerable with the adequate set of harmonics. 
 
 \vspace{-0.3cm}
\subsection{Gaussian Background Models} \label{app:randomModels}	
	
	{ A two-dimensional  random field $S : \mathbb{R}^2 \rightarrow \mathbb{R}$ can be observed through a test function $f$, in which case the quantity 
	$$\langle S , f \rangle = \int_{\R^2} S(\bm{x}) f(\bm{x}) \mathrm{d}\bm{x}$$
	is a random variable. 
	We say that $S$ is Gaussian if all the random variables $\langle S , f \rangle$ are Gaussian for any test function.}
	
	The Gaussian white noise   $W$ is probably the most famous two-dimensional Gaussian field. It is the continuous-domain generalization of a family of independent and identically distributed Gaussian random variables (discrete Gaussian white noise). 
	The white noise $W$ is stationary and such that $\langle W , f \rangle$ and $\langle W , g \rangle$ are independent as soon as $\langle f,g \rangle = 0$.
	Observing $W$ through a test function $f \in L_2(\R^2)$ gives a Gaussian random variable 	$\langle W , f \rangle$ with zero-mean and variance  $\sigma^2 \lVert f \rVert^2_2$~\cite{Unser2014sparse}.	
	We then call $\sigma^2$ the variance of $W$.

	One can more generally consider random fields $S$ such that $\Lop \{S\} = W$ is a Gaussian white noise, where $\Lop$ is a linear differential operator. 
	As developed more extensively in \cite{Unser2014sparse}, one can deduce the expression of the variance of $\langle S , f \rangle$ from the variance of the white noise as follows. 
	Consider a function $g \in L_2(\R^2)$ and set $f = \Lop^* \{g\}$, with $\Lop^{*}$ the adjoint of $\Lop$. 
	Then, we have by duality that
\begin{equation}\label{eq:trucvariance}
	\langle S , f \rangle = \langle S , \Lop^* \{g\}\rangle = \langle \Lop \{S\} , g \rangle = \langle W , g \rangle. 
\end{equation}
In particular, we deduce that $\langle S , f \rangle \sim \mathcal{N} ( 0 , \sigma^2 \lVert g \rVert_2^2 )$, which gives \eqref{eq:varianceS}. 

{In our case, we select $\Lop = (-\Delta)^{\gamma/2}$ {to be the fractional Laplacian~\cite{Lodhia2016fractional}}, which is self-adjoint in the sense that $((-\Delta)^{\gamma/2})^* = (-\Delta)^{\gamma/2}$. 
This corresponds to the power spectrum $P_S(\bm{\omega}) = \sigma^2 / r^{2\gamma}$ as in Section \ref{subsec:spectralshaping}.}
We then  have    $\widehat{f}(\bm{\omega}) = \lVert \bm{\omega} \rVert^\gamma \widehat{g}(\bm{\omega})$. Hence,
\begin{align}
	\lVert {g} \rVert_2^2 &= \frac{1}{2\pi} \lVert \widehat{g} \rVert_2^2  = \frac{1}{2\pi} \int_{\R^2} \frac{|\widehat{f}(\bm{\omega})|^2}{ \lVert \bm{\omega} \rVert^{2\gamma}} \mathrm{d}\bm{\omega} \nonumber \\
	&= \frac{1}{2\pi} \int_0^{\infty} r^{1-2\gamma} \int_0^{2\pi} \lvert \widehat{f}(r,\theta) \rvert^2 \mathrm{d}\theta \mathrm{d}r
\end{align}
implying \eqref{eq:varianceS}.

\vspace{-0.3cm}
\subsection{Optimal Steerable Filter: With Spectral Shaping}
\label{ProofSpectralShaping}

The main idea is to reduce Proposition \ref{prop:spectralshaping} to Theorem \ref{theo:optimalfilter}.
The criterion \eqref{eq:SNR3} is optimized among the filters $f = (-\Delta)^{\gamma/2} g$ for some $g \in L_2(\R^2)$. 
For such filters,  we have $\widehat{f} = \lVert \cdot \rVert^{\gamma} \widehat{g}$, implying that
\begin{equation}
	\frac{\rvert\langle \widehat{T} , \widehat{f} \rangle\lvert^2}{\lVert \widehat{g}  \rVert_2^2} =
	\frac{\rvert\langle  \widehat{T} ,  \lVert \cdot \rVert^{\gamma} \widehat{g} \rangle\lvert^2}{\lVert \widehat{g} \rVert_2^2}
	=	
	\frac{\rvert\langle \lVert \cdot \rVert^{\gamma} \widehat{T} , \widehat{g} \rangle\lvert^2}{\lVert \widehat{g} \rVert_2^2}.
\end{equation}
Therefore, $f$ maximizes \eqref{eq:SNR3} if and only if $g$  maximizes $\frac{\rvert\langle  \lVert \cdot \rVert^{\gamma} \widehat{T} , \widehat{g} \rangle\lvert^2}{\lVert \widehat{g} \rVert_2^2}$ among square integrable functions. The latter is equivalent, according to Theorem \ref{theo:optimalfilter}, to $\widehat{g}(r,\theta) = \sum_{n \in H} {(\lVert \cdot \rVert^{\gamma}\widehat{T})}_n (r) \mathrm{e}^{\mathrm{j} n \theta}$ where the ${(\lVert \cdot \rVert^{\gamma}\widehat{T})}_n$ are the Fourier radial profiles of $\lVert \cdot \rVert^{\gamma}\widehat{T}$. Because $\lVert \cdot \rVert^{\gamma}$ is isotropic, we easily get that
${(\lVert \cdot \rVert^{\gamma}\widehat{T})}_n (r) =  r^\gamma \widehat{T}_n(r)$. Finally,   $f$ maximizes \eqref{eq:SNR3} if and only if 
\begin{align}
	\widehat{f}(r,\theta) \propto r^\gamma \sum_{n \in H}  r^\gamma \widehat{T}_n(r) \mathrm{e}^{\mathrm{j} n \theta}
	 = r^{2\gamma} \sum_{n \in H}  \widehat{T}_n(r) \mathrm{e}^{\mathrm{j} n \theta}
\end{align}
and \eqref{eq:fSNR2} is proved.

\vspace{-0.3cm}
\subsection{Computing Radial B-spline Expansions}
\label{app:Bsplineprop}

We here provide the proofs of Proposition \ref{prop:radialexpansion} and Theorem \ref{theo:Bsplineexpansion}.
In the two cases, the main argument is the following classical result, that can be found for instance in \cite{Aldroubi1994sampling}. \\

\vspace{-0.3cm}
\begin{proposition} \label{prop:abstractgramprojection}
Assume that $(\varphi_k)_{k \in \Z}$ is a family of square integrable functions forming a Riesz basis; that is, satisfying 
\begin{equation}
	A \sum_{n \in \Z} c[k]^2 \leq \big\lVert \sum_{k \in \Z} c[k] \varphi_k \big\rVert_{2}^2 \leq B  \sum_{k \in \Z} c[k]^2,
\end{equation}
with $0<A\leq B< \infty$.
Then, the orthogonal projection onto the span $V$ of the $\varphi_k$ is of the form
\begin{equation}
	\mathrm{P}_V\{ f \} = \sum_{k \in \Z} c[k] \varphi_k,
\end{equation} 
where {  the sequence $c = (c[k])$ satisfies the relation
\begin{equation}  \label{eq:Gcd}
 G\{c \} [k] = \sum_{\ell\in \Z} G[k,\ell] c[\ell] = d[k],
\end{equation}
for every $k \in \Z$. The infinite matrix $G = (G[k,\ell] )$ and the sequence $d=(d[k])$ are defined as
\begin{equation}
	G[k,\ell] = \langle \varphi_k , \varphi_\ell \rangle \ \text{ and } \ d[k] = \langle f , \varphi_k \rangle.
\end{equation}}
\end{proposition}
\vspace{-0.3cm}

{ The (infinite) matrix $G$ is called the Gram matrix. The relation \eqref{eq:Gcd} can be compacted as $G\{c\} = d$, which we shall use thereafter.}
Proposition \ref{prop:abstractgramprojection} allows to express the expansion of functions in non-orthonormal basis from the quantities $\langle f , \varphi_k \rangle$. 

We can now prove Proposition \ref{prop:radialexpansion}. 
The family $\left(\frac{1}{r_0} \beta ( \cdot / r_0 - k)\right)_{k \in \mathbb{Z}}$ forms a Riesz basis of $L_2(\R)$ \cite{Unser1999splines}, and this property is easily extended to the case of radial functions of $L_2(\R^2)$. 
Therefore, the coefficients $d[k]$ are  given by \eqref{eq:thedk} and the Gram matrix satisfies
\begin{align}
	G [k,\ell] &= \left\langle \frac{1}{r_0} \beta\left( \frac{r}{r_0} - k\right) ,  \frac{1}{r_0} \beta\left( \frac{r}{r_0} - \ell \right) \right\rangle \nonumber \\
	&=  \int_{\R} \beta\left( \frac{r}{r_0} - k\right) \beta\left( \frac{r}{r_0} -  \ell \right) \frac{r}{r_0} \mathrm{d} \left( \frac{r}{r_0} \right)\nonumber \\
	&= \int_{\R} \beta\left( {r} \right) \beta\left( {r} -  (\ell - k) \right) r \mathrm{d} r,
\end{align}
where we used the change of variable $r \leftarrow (r / r_0 - k)$. 
In particular, $G[k,\ell]$ does not depend on $r_0$ and only on the difference $(\ell-k)$. Denoting $g[k] = G[0,k]$, we hence have that $G \{ c \} = g * c = d$, which is equivalent to \eqref{eq:cviad} and proves Proposition \ref{prop:radialexpansion}. 

We obtain Theorem \ref{theo:Bsplineexpansion} with the same arguments applied to the family $(\varphi_{n,k})_{n\in H, k \in \Z}$. We recall that two functions $\varphi_{n,k}$ and $\varphi_{m,\ell}$ are orthogonal as soon as $n \neq m$ since the circular harmonics are. Therefore, one can treat the problem independently for each harmonics and apply Proposition \ref{prop:radialexpansion} on the Fourier radial profiles $\widehat{T}_n$ of ${T}$. 

For the last point, quadratic splines are known to well approximate functions from $\R$ to $\R$ up to an arbitrary precision \cite{Unser1999splines}. This fact is easily adapted to the case of two-dimensional radial functions.  We then deduce that, for each $n$, the orthogonal projection of $\widehat{T}_n$ converges to $\widehat{T}_n$ when the step size $r_0\rightarrow 0$. Then, we remark, using the triangular inequality and the orthogonal relations between circular harmonics, that

\small
\begin{align} \label{eq:controlTviaHr0}
	\lVert \widehat{T} -  \mathrm{P}_{H,r_0} & \{ \widehat{T} \} \rVert_2 
		 \leq 
		\lVert \widehat{T} - \mathrm{P}_{H} \{ \widehat{T} \} \rVert_2  + \lVert  \mathrm{P}_{H} \{ \widehat{T} \} - \mathrm{P}_{H,r_0} \{ \widehat{T} \} \rVert_2 \nonumber \\
		&=
		\lVert \widehat{T} - \mathrm{P}_{H} \{ \widehat{T} \} \rVert_2 + \left( \sum_{n \in H} \lVert \widehat{T}_n - \mathrm{P}_{r_0} \{ \widehat{T}_n \} \rVert_2^2\right)^{1/2}
\end{align}
\normalsize
where $\mathrm{P}_{r_0}$ is given in \eqref{eq:projectionBsplines}. 
For $H$ finite and large enough, the first quantity in \eqref{eq:controlTviaHr0} is arbitrarily small. Then, for such $H$, one can select $r_0$ such that $\lVert \widehat{T}_n - \mathrm{P}_{r_0} \{ \widehat{T}_n \} \rVert_2$ is arbitrarily small for each $n \in H$. Finally, one can approximate $\widehat{T}$ with arbitrary precision for $H$ large enough and $r_0$ small enough. 

\section*{Acknowledgements}

The authors would like to thank Emrah Bostan for his precious help regarding the statistical estimation of self-similarity parameters. They also indebted to Daniel Sage for fruitful discussions on the manuscript. \review{Finally, the authors are sincerely grateful to Thierry Blu and Tianle Zhao for sharing their implementation of Fourier-Argand steerable filters, allowing us to perform fair comparative experiments. Their exemplary scientific openness greatly contributed to improve this paper.}

\bibliographystyle{ieeetr}
\bibliography{references}

\end{document}